\documentclass{emulateapj}
\usepackage{epsfig}
\usepackage{graphicx}
\usepackage{psfig,epsf}

\def\BF  {\mathbf F}
\def\dsi  {\mathbf {dS_i}}
\def\cdt   {{1 \over {c \Delta t}}}
\def\dJgdt {{1 \over c} {\partial J_g \over {\partial t}}}

\def\sun{_\odot}
\def\kms{\,km\,s$^{-1}$}
\def\mo{M$_\odot$\,}
\def\mdot{\,M$_\odot$\,s$^{-1}$\,}

\def\Dwa{$\,$\uppercase\expandafter{\romannumeral5}$\,$}

\def\sles{\lower2pt\hbox{$\buildrel {\scriptstyle <}
   \over {\scriptstyle\sim}$}}

\def\sgreat{\lower2pt\hbox{$\buildrel {\scriptstyle >}
   \over {\scriptstyle\sim}$}}
\def\sharpnull#1{}
\def\aa{Astron. Astrophys.\ }

\begin{document}

\title{Multi-Dimensional Radiation/Hydrodynamic Simulations of Protoneutron
Star Convection}

\author{L. Dessart\altaffilmark{1},
A. Burrows\altaffilmark{1},
E. Livne\altaffilmark{2},
C.D. Ott\altaffilmark{3}}
\altaffiltext{1}{Department of Astronomy and Steward Observatory,
                 The University of Arizona, Tucson, AZ \ 85721;
                 luc@as.arizona.edu,burrows@as.arizona.edu}
\altaffiltext{2}{Racah Institute of Physics, The Hebrew University,
Jerusalem, Israel; eli@frodo.fiz.huji.ac.il}
\altaffiltext{3}{Max-Planck-Institut f\"{u}r Gravitationsphysik,
Albert-Einstein-Institut, Golm/Potsdam, Germany; cott@aei.mpg.de}

\begin{abstract}

Based on multi-dimensional, multi-group, flux-limited-diffusion 
hydrodynamic simulations of core-collapse supernovae
with the VULCAN/2D code, we study the physical conditions within and in the vicinity of the
nascent protoneutron star (PNS).
Our numerical study follows the evolution of the collapsing envelope of the 11-M$_{\odot}$ model
of Woosley \& Weaver, from $\sim$200\,ms before bounce to $\sim$300\,ms after bounce, 
on a spatial grid that switches from Cartesian at the PNS center to spherical 
above a 10\,km radius.

As has been shown previously, we do not see any large-scale overturn of the inner PNS material. 
Convection, directly connected to the PNS, is found to occur in two distinct regions: between
10 and 20\,km, coincident with the region of negative lepton gradient, and
exterior to the PNS above 50\,km. Separating these two regions, an interface, 
with no sizable outward or inward motion, is the site of gravity waves, 
emerging at $\sim$200-300\,milliseconds (ms) after 
core bounce, excited by the convection in the outer convective zone.

In the PNS, convection is always confined within the neutrinospheric radii for 
all neutrino energies above just a few MeV. We find that such convective motions do not 
appreciably enhance the $\nu_e$ neutrino luminosity, 
and that they enhance the $\bar{\nu}_e$ and ``$\nu_{\mu}$" luminosities modestly,
by $\sim$15\% and $\sim$30\%, respectively, during the first post-bounce 100--200\,ms.
Moreover, we see no evidence of doubly-diffusive instabilities in the PNS, expected to 
operate on diffusion timescales of at least a second, much longer than the millisecond
timescale associated with PNS convection.

PNS convection is thus found to be a secondary feature of the core-collapse 
phenomenon, rather than a decisive ingredient for a successful explosion.

\end{abstract}


\section{Introduction}
\label{intro}

A few million years after the onset of core-hydrogen burning on the main
sequence, a massive star creates a degenerate core, which, upon reaching the
Chandrasekhar mass, undergoes gravitational collapse.
The core infall is eventually halted when its innermost regions reach nuclear densities,
generating a shock wave that propagates outwards, initially reversing inflow into outflow.
However, detailed numerical radiation-hydrodynamics simulations struggle to produce explosions.
Instead of a prompt explosion occurring on a dynamical timescale, simulations
produce a stalled shock at $\sim$100-200\,km, just a few tens of milliseconds
after core bounce: 1D simulations universally yield a stalled shock that ultimately
fails to lift off; 2D simulations provide a
more divided picture, with success or failure seemingly dependent on the numerical
approach or physical assumptions made.
Indeed, energy deposition by neutrinos behind the shock, in the so-called gain region,
is expected to play a central role in re-energizing the stalled shock:
explosion will occur if the shock can be maintained at large-enough radii
and for a sufficiently long time to eventually expand into the 
tenuous infalling envelope (Burrows \& Goshy 1993).
At present, the failure to produce explosions may be due to physical processes not 
accounted for (e.g., magnetic fields), to an inaccurate treatment of 
neutrino transport (see the discussion in Buras et al. 2005),
or to missing neutrino microphysics.

Because neutrino transport is a key component of the supernovae mechanism in
massive stars, slight modifications in the neutrino cooling/heating efficiency
could be enough to produce an explosion.
Alternatively, an enhancement in the neutrino flux during the first second
following core bounce could also lead to a successful explosion.
Convection in the nascent protoneutron star (PNS) has been invoked as a potential
mechanism for such an increase in the neutrino luminosity.
Neutrino escape at and above the neutrinosphere establishes a negative lepton-gradient, a
situation unstable to convection under the Ledoux (1947) criterion.
Epstein (1979) argued, based on a simulation by Mayle \& Wilson (1988), that this leads to large-scale 
overturn and advection of core regions with high neutrino-energy density at and 
beyond the neutrinosphere, thereby enhancing the neutrino flux.
Such large scale overturn was obtained in simulations by Bruenn,
Buchler \& Livio (1979) and Livio, Buchler \& Colgate (1979), but, as shown by
Smarr et al. (1981), their results were compromised by an inadequate equation of state (EOS).
Lattimer \& Mazurek (1981) challenged the idea of large-scale overturn, noting
the presence of a positive, stabilizing entropy gradient - a residue of the shock birth.
Thus, while large-scale overturn of the core is unlikely and has thus far
never been seen in realistic simulations of core-collapse supernovae, the possibility
of convectively-enhanced neutrino luminosities was still open.
Burrows (1987), based on a simple mixing-length treatment, argued that
large-scale core-overturn was unlikely, but found neutrino-luminosity enhancements, 
stemming from convective motions within the PNS, of up to 50\%.
Subsequently, based on multi-dimensional (2D) radiation hydrodynamics simulations, Burrows, 
Hayes, \& Fryxell (1995) did not find large-scale core overturn, nor any luminosity enhancement,
but clearly identified the presence of convection within the PNS, as well as ``tsunamis'' 
propagating at its surface.
Keil, Janka, \& M\"{u}ller (1996), using a similar approach, reported the presence of both convection
and enhancement of neutrino luminosities compared to an equivalent 1D configuration.
However, these studies used gray neutrino transport,  
a spherically-symmetric (1D) description of the
inner core (to limit the Courant timestep), and a restriction of the angular coverage to 90$^{\circ}$.
Keil et al. also introduced a diffusive and moving outer boundary at the surface of the PNS (receding 
from 60\,km down to 20\,km, 1\,s after core bounce), thereby neglecting any feedback from the 
fierce convection occurring above, between the PNS surface and the shock radius.
Mezzacappa et al. (1998) have performed 2D hydrodynamic simulations of
protoneutron star convection, with the option of including 
neutrino-transport effects as computed from equivalent 1D MGFLD
simulations. They found that PNS convection obtains underneath the neutrinosphere
in pure hydrodynamical simulations, but that neutrino transport considerably
reduces the convection growth rate. We demonstrate in the present work 
that PNS convection does in fact obtain, even with a (multidimensional) 
treatment of neutrino transport and that, as these authors anticipated, 
the likely cause is their adopted 1D radiation transport, which maximizes 
the lateral equilibration of entropy and lepton number.
Recently, Swesty \& Myra (2005ab) described simulations of the convective epoch in core-collapse 
supernovae, using MGFLD neutrino transport, but their 2D study covers only the initial
33\,milliseconds (ms) of PNS evolution.

Alternatively, Mayle \& Wilson (1988) and Wilson \& Mayle (1993) have argued
that regions stable to convection according to the Ledoux criterion could be the sites 
of doubly-diffusive instabilities, taking the form of so-called neutron (low-$Y_{\rm e}$ material) fingers.
This idea rests essentially on the assumption that the neutrino-mediated diffusion of heat
occurs on shorter timescales than the neutrino-mediated diffusion of leptons.
By contrast, Bruenn \& Dineva (1996) and Bruenn, Raley, \& Mezzacappa (2005) demonstrated
that the neutrino-mediated thermal-diffusion timescale is longer than that of
neutrino-mediated lepton-diffusion, and, thus, that neutron fingers do not obtain.
Because  $\nu_{\mu}$'s and $\nu_{\tau}$'s have a weak thermal coupling
to the material, Bruenn, Raley, \& Mezzacappa (2005) concluded that lepton-diffusion would occur faster
by means of low-energy ${\bar{\nu}}_{\rm e}$'s and $\nu_{\rm e}$'s.
Applying their transport simulations to snapshots of realistic core-collapse simulations,
they identified the potential for two new types of instabilities within the PNS, referred 
to as ``lepto-entropy fingers" and ``lepto-entropy semiconvection".

In this paper, we present a set of simulations that allow a consistent assessment
of dynamical/diffusive/convective mechanisms taking place within the PNS, improving on a number of
assumptions made by previous radiation-hydrodynamic investigations.
Our  approach, based on VULCAN/2D (Livne et al. 2004), has several
desirable features for the study of PNS convection.   
First, the 2D evolution of the inner 3000-4000\,km of the core-collapsing massive star
is followed from pre-bounce to post-bounce. Unlike other groups (Janka \&
M\"{u}ller 1996; Swesty \& Myra 2005ab), for greater consistency we do not 
start the simulation at post-bounce times by remapping a 1D simulation evolved till core bounce.
Second, the VULCAN/2D grid, by switching from Cartesian (cylindrical) in the inner region (roughly
the central 100\,km$^2$) to spherical above a few tens of kilometers (chosen as desired), 
allows us to maintain 
good resolution, while preventing the Courant timestep from becoming prohibitively small. 
Unlike previous studies of PNS convection (e.g., Swesty \& Myra 2005ab),  
we extend the grid right down to the center, without the excision of the inner kilometers.
Additionally, this grid naturally permits the core to move and, thus, in principle, 
provides a consistent means to assess the core recoil associated with asymmetric explosions.
Third, the large radial extent of the simulation, from the inner core to a few thousand
km, allows us to consider the feedback effects between different regions.
Fourth, our use of Multi-Group Flux-Limited Diffusion is also particularly suited for
the analysis of mechanisms occurring within a radius of 50\,km, since
there, neutrinos have a diffusive behavior, enforced by the high opacity of the
medium at densities above 10$^{11}$\,g\,cm$^{-3}$. 
Fifth, lateral transport of neutrino energy and lepton number is accounted 
for explicitly, an asset over the more approximate ray-by-ray approach 
(Burrows, Hayes, \& Fryxell 1995; Buras et al. 2005) which cannot 
simulate accurately the behavior of doubly-diffusive instabilities.
A limitation of our work is the neglect of the subdominant inelastic e$^{\rm -}-\nu_{\rm e}$ 
scattering in this version of VULCAN/2D.

The present paper is structured as follows. In \S\ref{code}, we discuss 
the VULCAN/2D code on which all simulations studied here are based.
In \S\ref{model}, we describe in detail the properties of our baseline model simulation,
emphasizing the presence/absence of convection, and limiting the discussion to the inner 50-100\,km.
In \S\ref{pns}, focusing on results from our baseline model, we characterize the PNS convection and
report the lack of doubly-diffusive instabilities within the PNS.
Additionally, we report the unambiguous presence of gravity waves, persisting over a few hundred 
milliseconds, close to the minimum in the electron-fraction distribution, at $\sim$20--30\,km.
In \S\ref{conclusion}, we conclude and discuss the broader significance of our results
in the context of the mechanism of core-collapse supernovae.

\section{VULCAN/2D and simulation characteristics}
\label{code}

   All radiation-hydrodynamics simulations presented in this paper were performed with
a time-explicit variant of VULCAN/2D (Livne 1993), adapted to model the mechanism 
of core-collapse supernovae (Livne et al. 2004;
Walder et al. 2005; Ott et al. 2004). The code uses cylindrical coordinates 
($r,z$) where $z$ is the distance parallel to the axis of symmetry (and, 
sometimes, the axis of  rotation) and $r$ is the position perpendicular to it
\footnote{The spherical radius, $R$, is given by the quantity $\sqrt{r^2 + z^2}$.}.
VULCAN/2D has the ability to switch from a Cartesian grid near the base to a 
spherical-polar grid above a specified transition radius $R_{\rm t}$. 
A reasonable choice is $R_{\rm t}\sim$20\,km since it allows
a moderate and quite uniform resolution of both the inner and the outer regions,
with a reasonably small total number of zones that extends out to a few thousand 
kilometers. With this setup, the ``horns,''\footnote{For 
a display of the grid morphology, see Fig.~4 in Ott et al. (2004).} associated with 
the Cartesian-to-spherical-polar transition, lie in the region where PNS convection obtains
and act as (additional) seeds for PNS convection (see below). 
We have experimented with alternate grid setups that place the horns either interior
or exterior to the region where PNS convection typically obtains, i.e., roughly between 10--30\,km
(Keil et al. 1996; Buras et al. 2005). We have performed three runs covering from
200\,ms before to $\sim$300\,ms after core bounce, using 25, 31, and 51 zones out to the 
transition radius at 10, 30, and 80\,km, placing the horns at 7, 26, and 65\,km, respectively. 
Outside of the Cartesian mesh, we employ 101, 121, and 141 angular zones, equally spaced over 180$^{\circ}$,
and allocate 141, 162, and 121  logarithmically-spaced zones between the transition radius and 
the outer radius at 3000, 3800, and 3000\,km, respectively.
The model with the transition radius at 30\,km modifies somewhat the timing of the appearance
of PNS convection; the model with the transition radius at 80\,km is of low resolution
and could not capture the convective patterns in the inner grid regions.
In this work, we thus report results for the model with the transition
radius at 10\,km, which possesses a smooth grid structure in the region where PNS convection 
obtains and a very high resolution of 0.25\,km within the transition radius, but has the 
disadvantage that it causes the Courant timestep to typically be a third of its value 
in our standard computations (e.g. Burrows et al. 2006), i.e., $\sim$3$\times$10$^{-7}$\,s.

As mentioned in \S\ref{intro}, this flexible grid possesses two assets: the grid resolution 
is essentially uniform everywhere interior to the transition radius, and, thus, 
does not impose a prohibitively small Courant timestep for our explicit hydrodynamic scheme, 
and the motion of the core is readily permitted, allowing estimates of potential core recoils 
resulting from global asymmetries in the fluid/neutrino momentum. 
The inner PNS can be studied right down to the core of the objects since no artificial 
inner (reflecting) boundary is placed there (Burrows, Hayes, \& Fryxell 1995; 
Janka \& M\"{u}ller 1996; Keil, Janka \& M\"{u}ller 1996; Swesty \& Myra 2005ab). 
Along the axis, we use a reflecting boundary, while at the outer grid radius, we
prevent any flow of material ($V_R = 0$), but allow the free-streaming of the neutrinos.

   We simulate neutrino transport using a diffusion approximation in 2D, together with 
a 2D version of Bruenn's (1985) 1D flux limiter; some details of our approach and
the numerical implementation in VULCAN/2D are presented in Appendix~A.
To improve over previous gray transport schemes, we solve the transport 
at different neutrino energies using a coarse, but satisfactory, sampling at 16 energy groups 
equally spaced in the log between 1 and 200\,MeV.
We have also investigated the consequences of using a lower energy resolution, with only 
8 energy groups, and for the PNS region we find no differences of a qualitative nature, 
and, surprisingly, few differences of a quantitative nature. 
While the neutrino energy distribution far above the neutrinosphere(s) (few 1000\,km) 
has a thermal-like shape with a peak at $\sim$15\,MeV and width of $\sim$10\,MeV, deep in 
the nascent PNS, the distribution peaks beyond 100\,MeV and is very broad.
In other words, we use a wide range of neutrino energies to solve the transport in order 
to model absorption/scattering/emission relevant at the low energies exterior to the 
neutrinospheres, and at the high energies interior to the neutrinospheres.

We employ the equation of state (EOS) of Shen et al. (1998), since it correctly
incorporates alpha particles and is more easily extended to lower densities
and higher entropies than the standard Lattimer \& Swesty (1991) EOS.
We interpolate in 180 logarithmically-spaced points in density, 180 logarithmically-spaced 
points in temperature, and 50 linearly-spaced points in electron fraction, 
whose limits are $\{\rho_{\rm min},\rho_{\rm max}\} = \{10^{5},10^{15}\}$\,(g\,cm$^{-3}$),
$\{T_{\rm min},T_{\rm max}\} = \{0.1,40\}$\,(MeV),
and $\{Y_{e,\rm min},Y_{e,\rm max}\} = \{0.05,0.513\}$.

The instabilities that develop in the early stages of the post-bounce
phase are seeded by noise at the part in $\sim$10$^6$ level in the EOS table interpolation.  
Beyond these, we introduce no artificial numerical perturbations. 
The ``horns'' associated with the Cartesian-to-spherical-polar transition 
are sites of artificially enhanced vorticity/divergence, with velocity magnitudes 
systematically larger by a few tens of percent compared to adjacent regions.
In the baseline model with the transition radius at 10\,km, this 
PNS convection sets in $\sim$100\,ms after core bounce, while in the simulation
with the transition radius (horns) at 30\,km (26\,km), it appears already quite developed
only $\sim$50\,ms after core bounce.
In that model, the electron-fraction distribution is also somewhat ``squared'' interior
to 20\,km, an effect we associate with the position of the horns, not present
in either of the alternate grid setups.
Thus, the Cartesian-to-spherical-polar transition introduces artificial seeds 
for PNS convection, although such differences are of only a quantitative
nature.
The most converged results are, thus, obtained with our baseline model, on which
we focus in the present work.
For completeness, we present, in Figs.~\ref{fig_entropy}--\ref{fig_density}, 
a sequence of stills depicting the pre-bounce evolution of our baseline model.

   In this investigation, we employ a single progenitor model, the 11\,\mo ZAMS model (s11) 
of Woosley \& Weaver (1995); when mapped onto our Eulerian grid, at the start of the simulation, 
the 1.33\,\mo Fe-core, which stretches out to 1300\,km, is already infalling.
Hence, even at the start of the simulation, the electron fraction extends from 0.5 above 
the Fe core down to 0.43 at the center of the object.

   Besides exploring the dependence on PNS convection of the number of energy groups, 
we have also investigated the effects of rotation.  For a model with an initial
inner rotational period of 10.47 seconds ($\Omega = 0.6$ rad s$^{-1}$), taken
from Walder et al. (2005), and with a PNS spin period of 10\,ms after $\sim$200\,ms 
(Ott et al. 2005), we see no substantive differences with our baseline model.
Hence, we have focused in this paper on the results from the non-rotating baseline model
\footnote{The consequences in the PNS core of much faster rotation rates will be 
the subject of a future paper.}. 

   Finally, before discussing the results for the baseline model, we emphasize that 
the word ``PNS'' is to be interpreted loosely: we mean by this the regions of the 
simulated domain that are within a radius of $\sim$50\,km. If the explosion fails, 
the entire progenitor mantle will eventually infall and contribute its mass to the 
compact object. If the explosion succeeds, it remains to be seen how much material 
will reach escape velocities.
At 300\,ms past core bounce, about 95\% of the progenitor core mass is within 30\,km.

\section{Description of the results of the baseline model}
\label{model}

\begin{figure}
\plotone{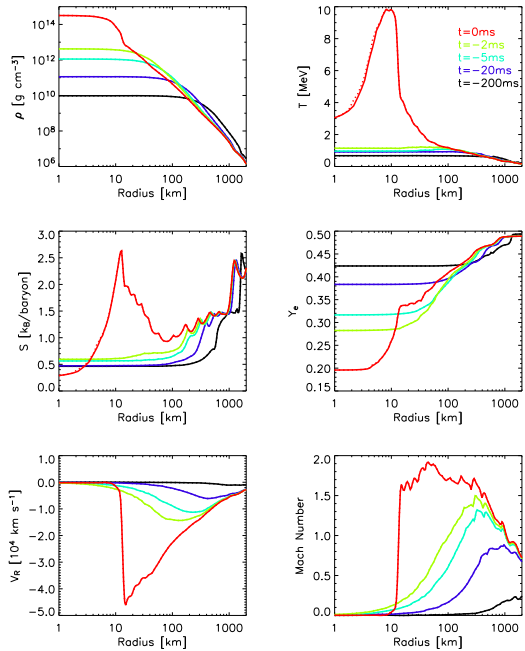}
\caption{
Montage of radial cuts along the polar (90$^{\circ}$; solid line) and equatorial (dotted line)
direction at 200 (black), 20 (blue), 5 (cyan), 2 (green), and 0\,ms (red) before core bounce
for the density (top left), temperature (top right), entropy (middle left), $Y_{\rm e}$ (middle right),
radial velocity (bottom left), and Mach number (bottom right).
}
\label{fig_pre_cc}
\end{figure}

\begin{figure*}
\plotone{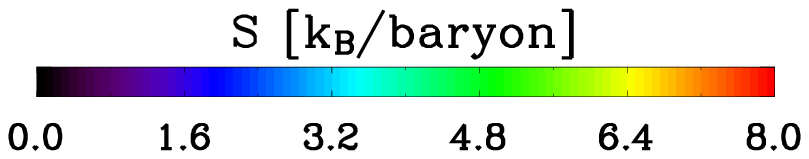}
\plottwo{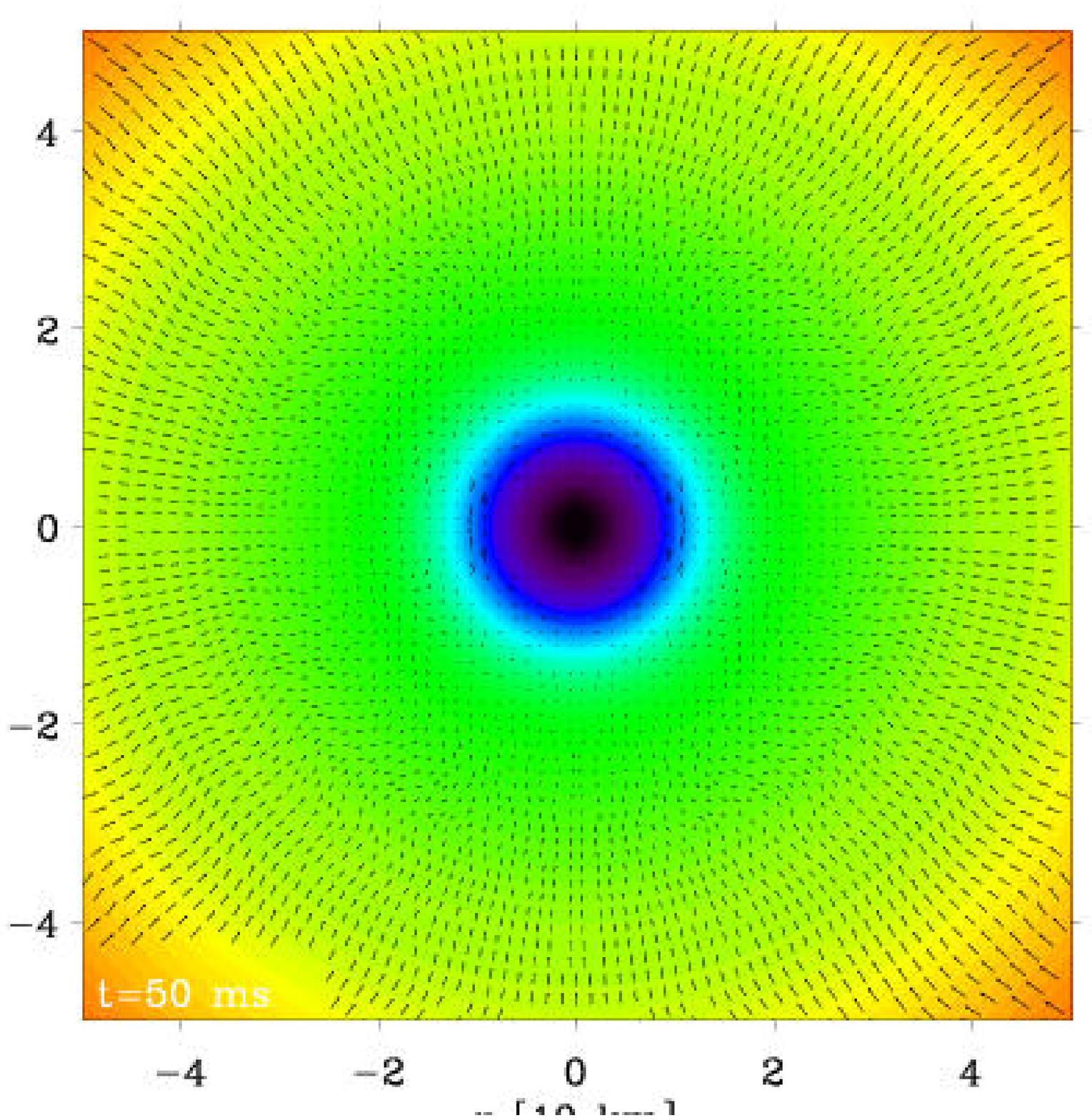}{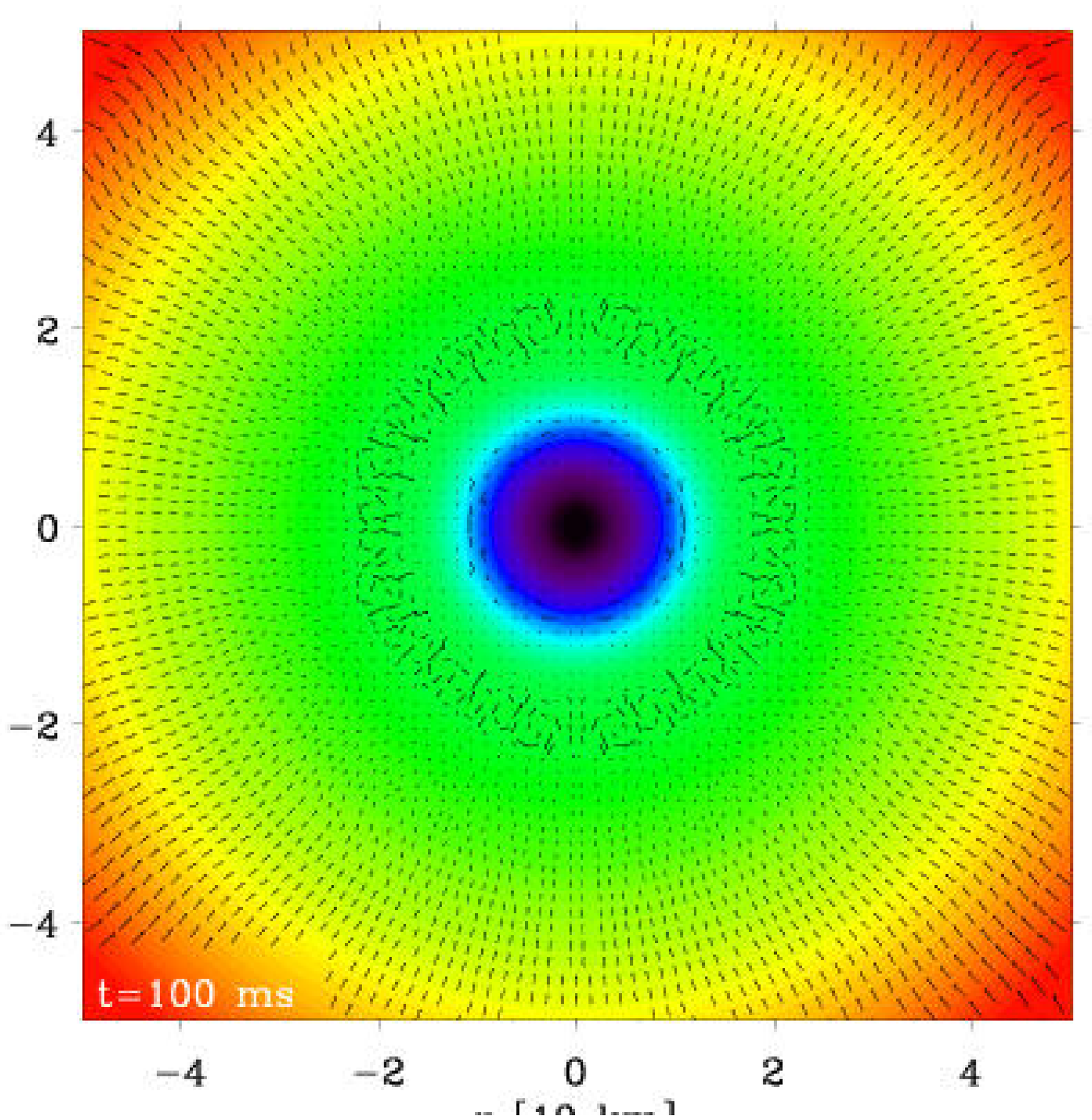}
\plottwo{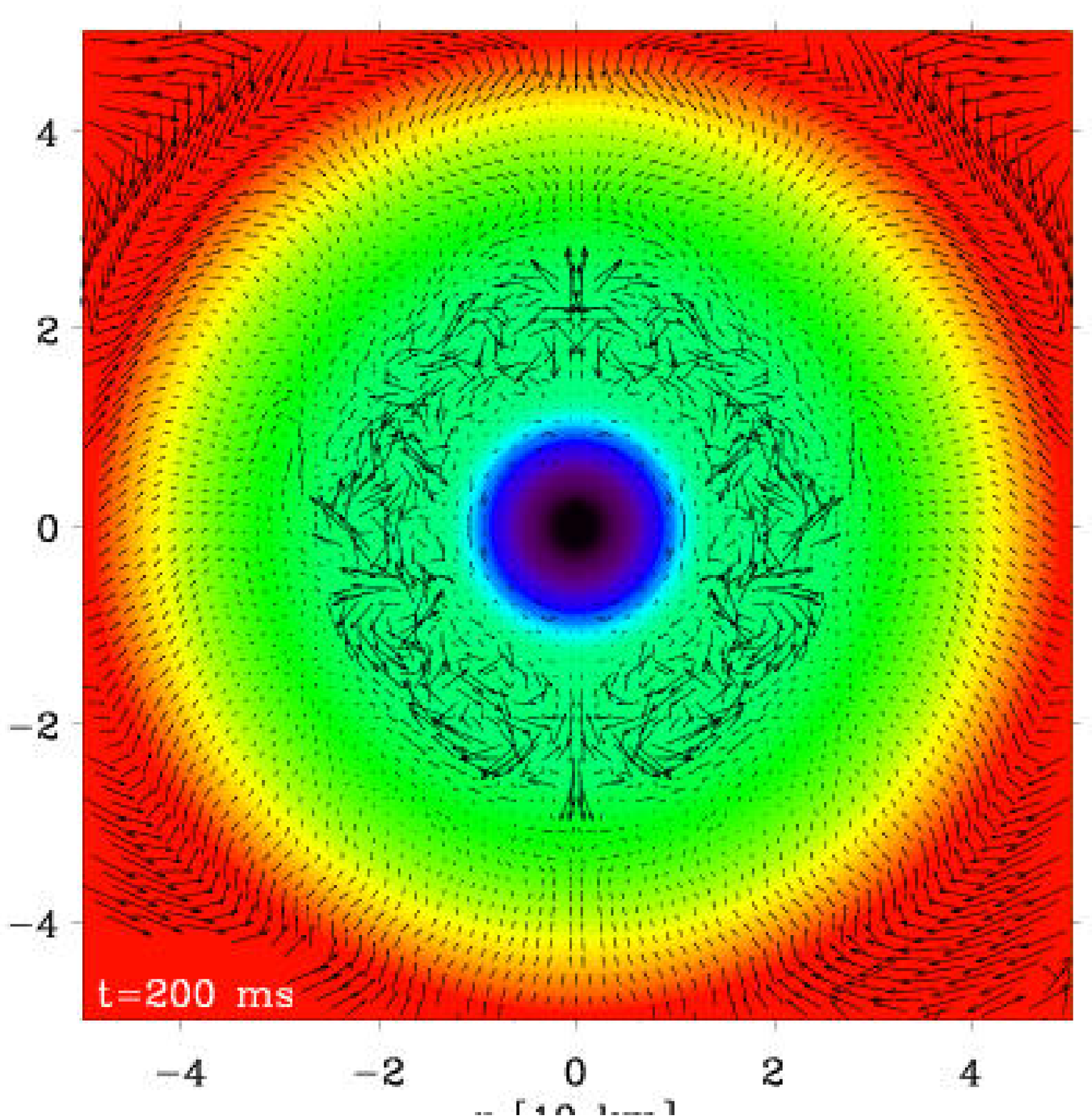}{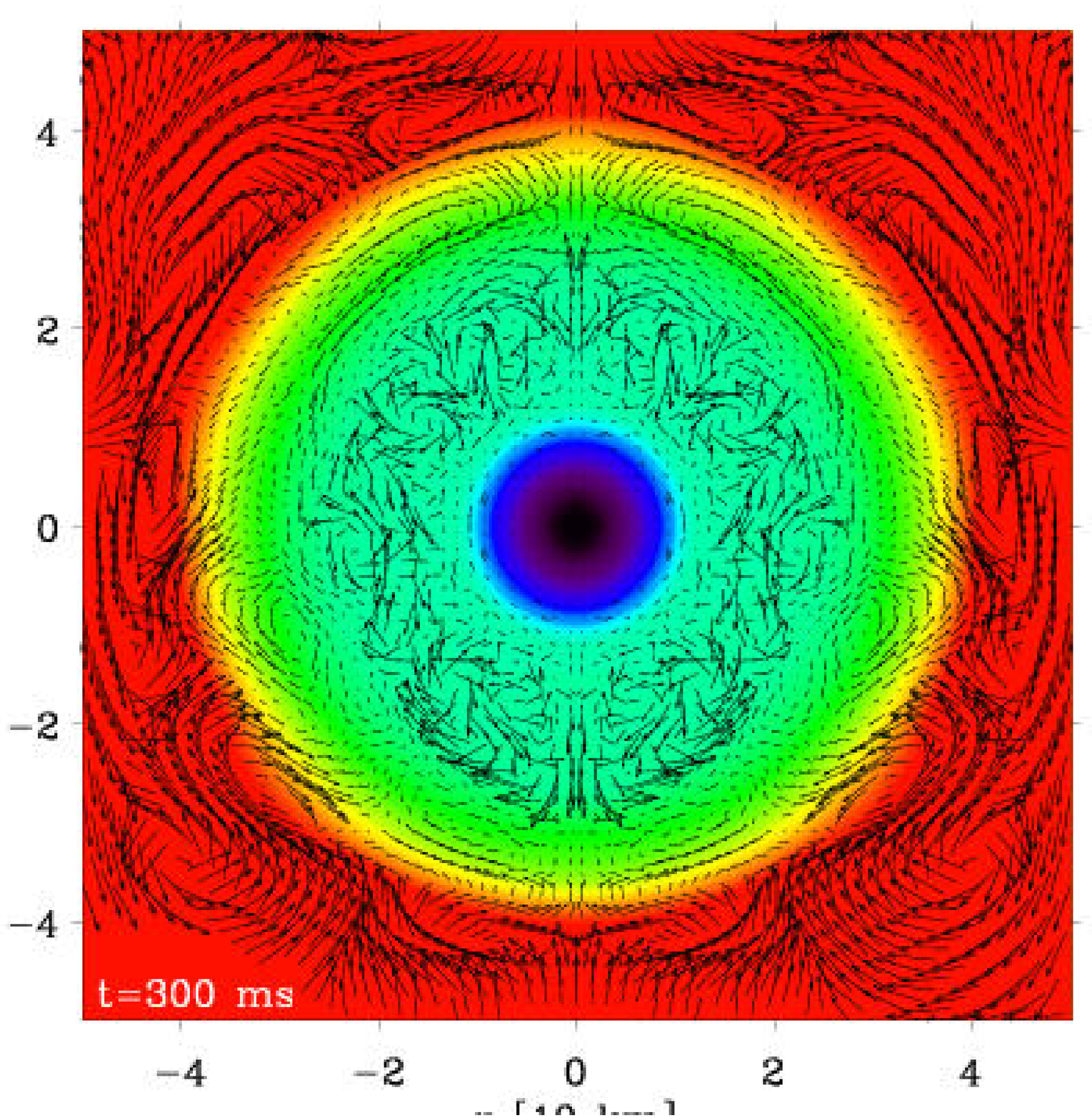}
\vspace{0.5cm}
\caption{
Color map stills of the entropy, taken at 50 (top left), 100 (top right), 
200 (bottom left), and 300\,ms (bottom right) past core bounce, with velocity vectors
overplotted.  Here ``Width" refers to the diameter; the radius through the middle is 50 kilometers.
Note that to ease the comparison between panels, the same range of values of the color map are used 
throughout (see text for discussion).
In all panels, the length of velocity vectors is saturated at 2000\,\kms, a value only
reached in the bottom-row panels. Note that the assessment of velocity magnitudes is best 
done using Fig.~\ref{fig_4slices} and Figs.~\ref{fig_DF_Vr}-\ref{fig_DF_Vt} (see text for discussion).
}
\label{fig_entropy}
\end{figure*}

\begin{figure*}
\plotone{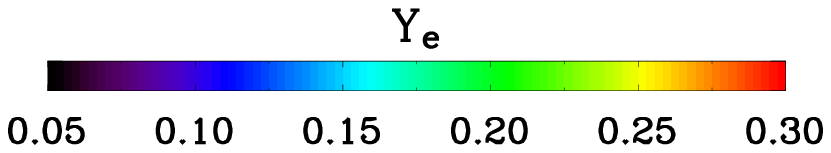}
\plottwo{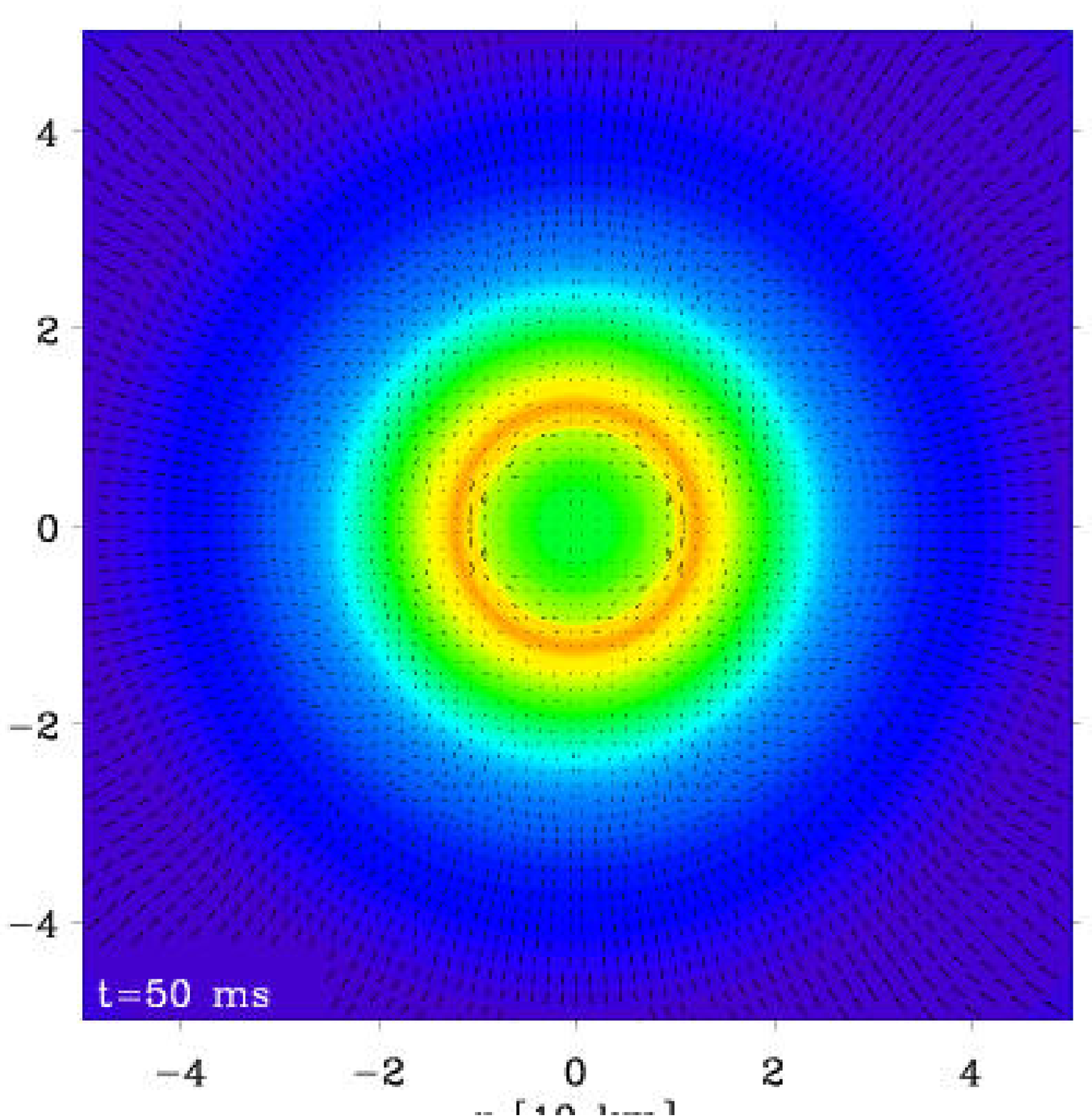}{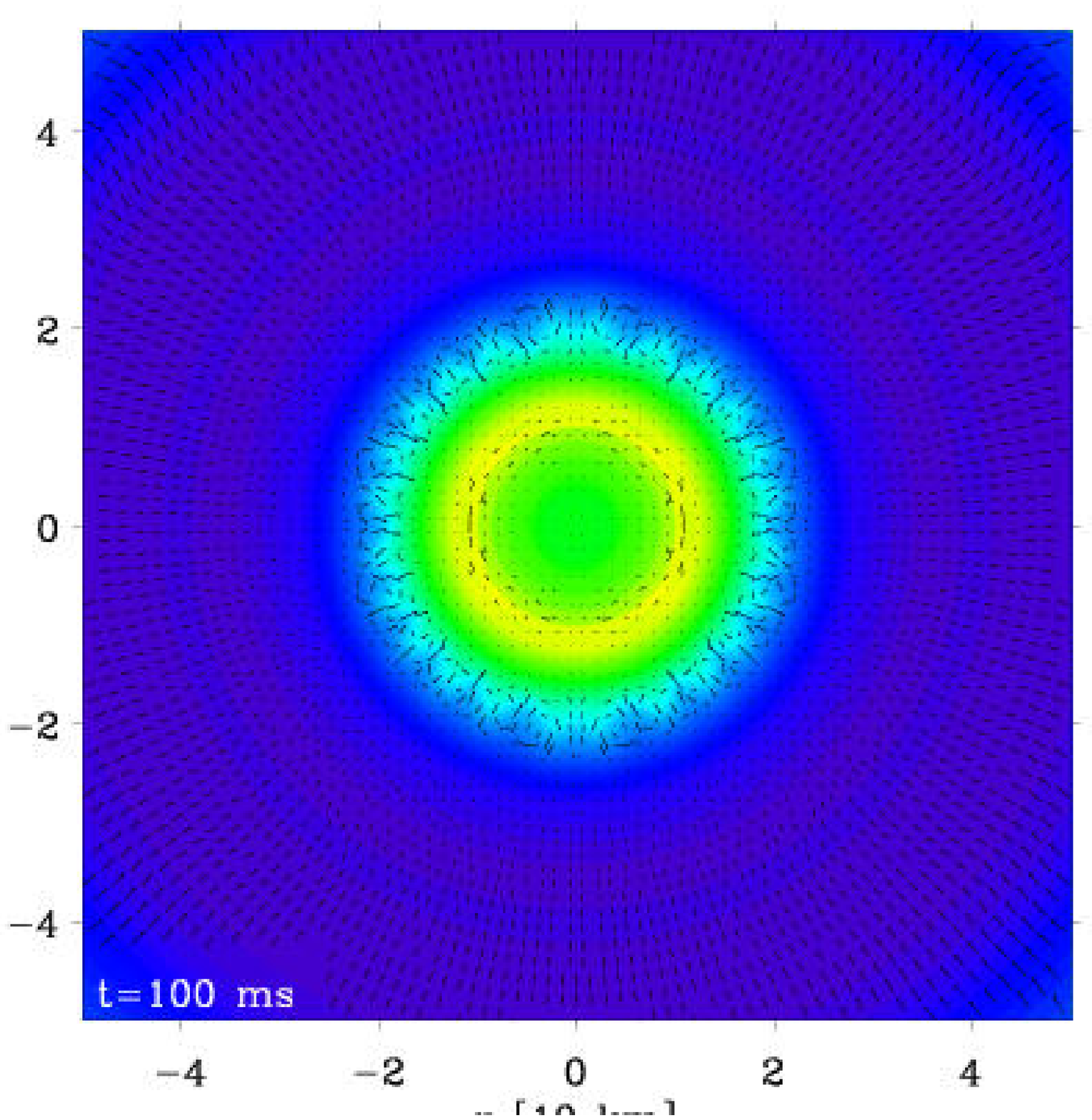}
\plottwo{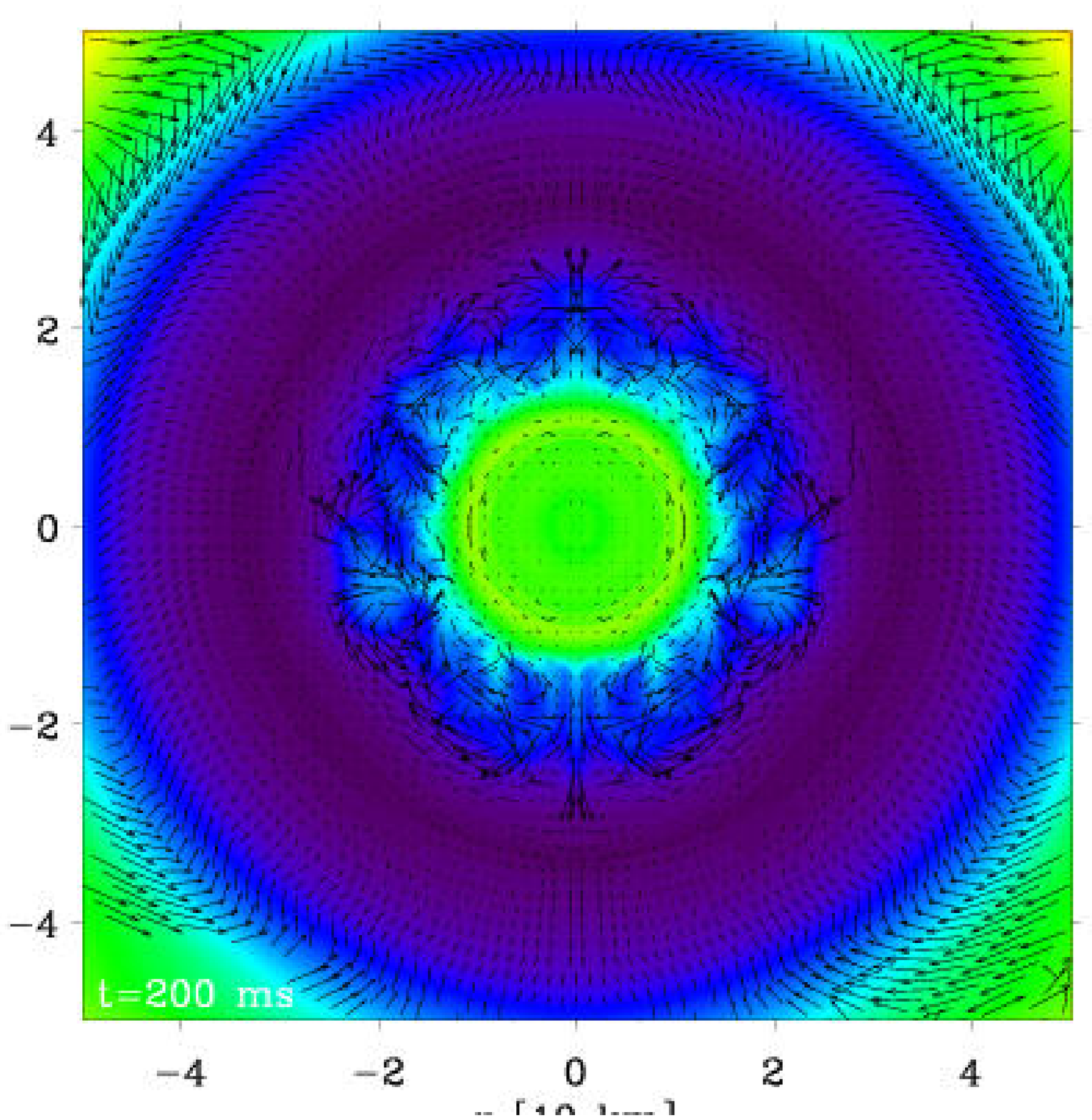}{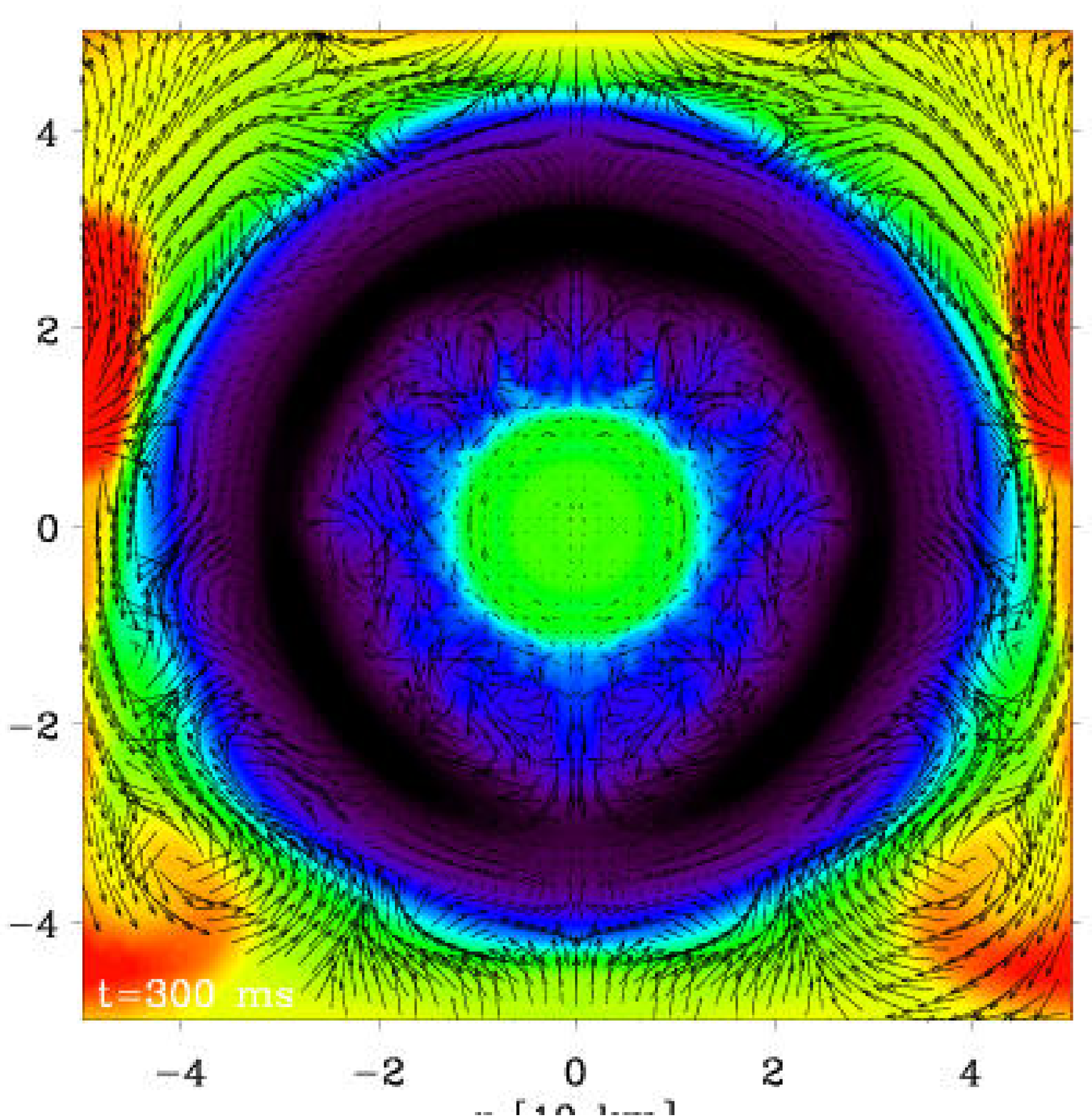}
\caption{
Same as Fig.~\ref{fig_entropy}, but for the electron fraction $Y_{\rm e}$.
(See text for discussion.)
}
\label{fig_ye}
\end{figure*}

\begin{figure*}
\plotone{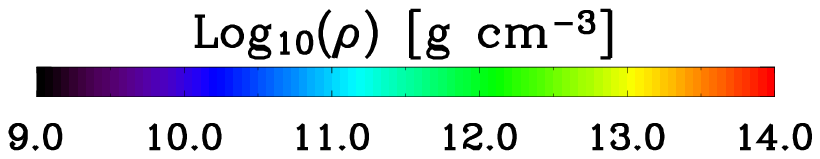}
\plottwo{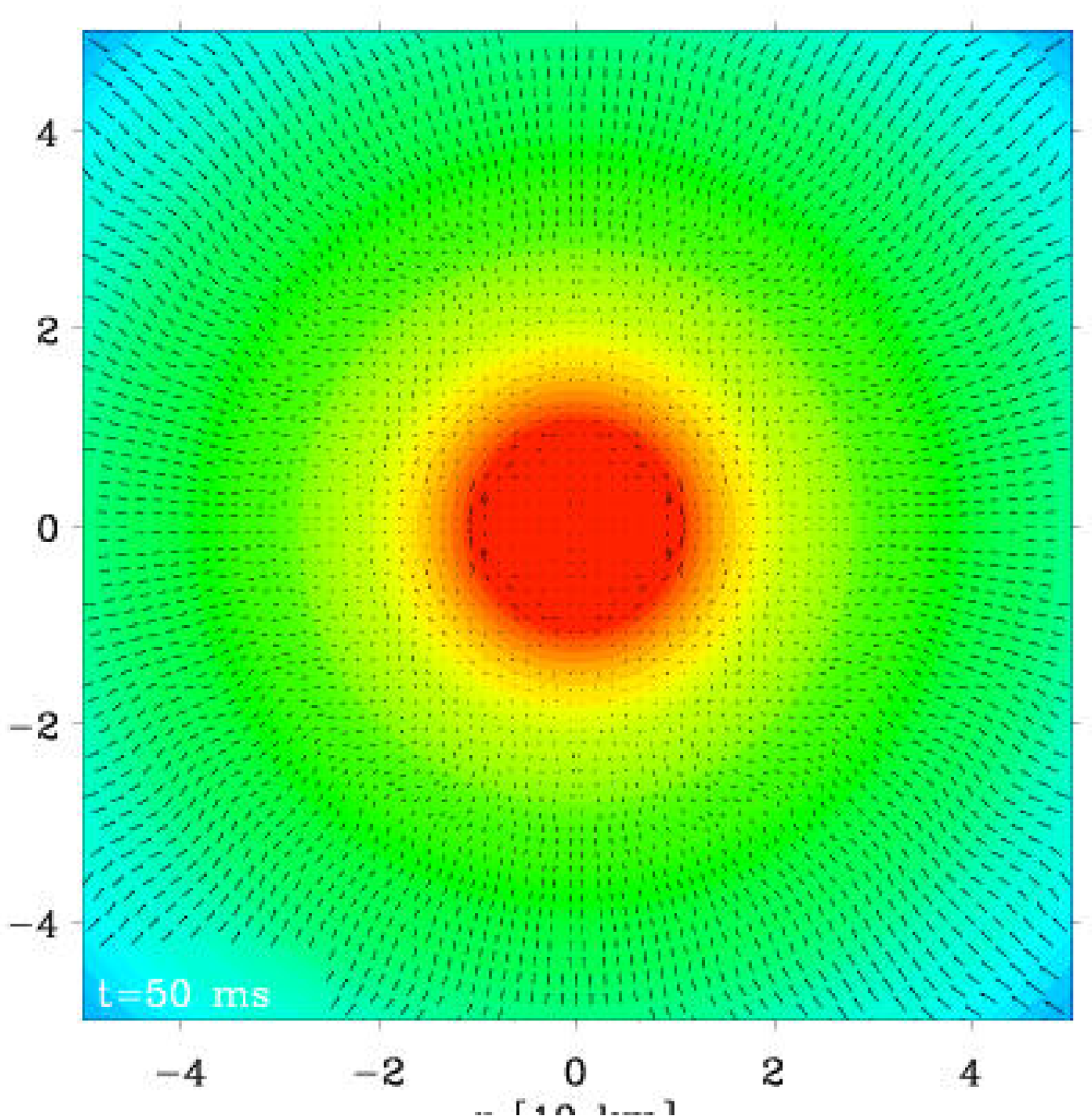}{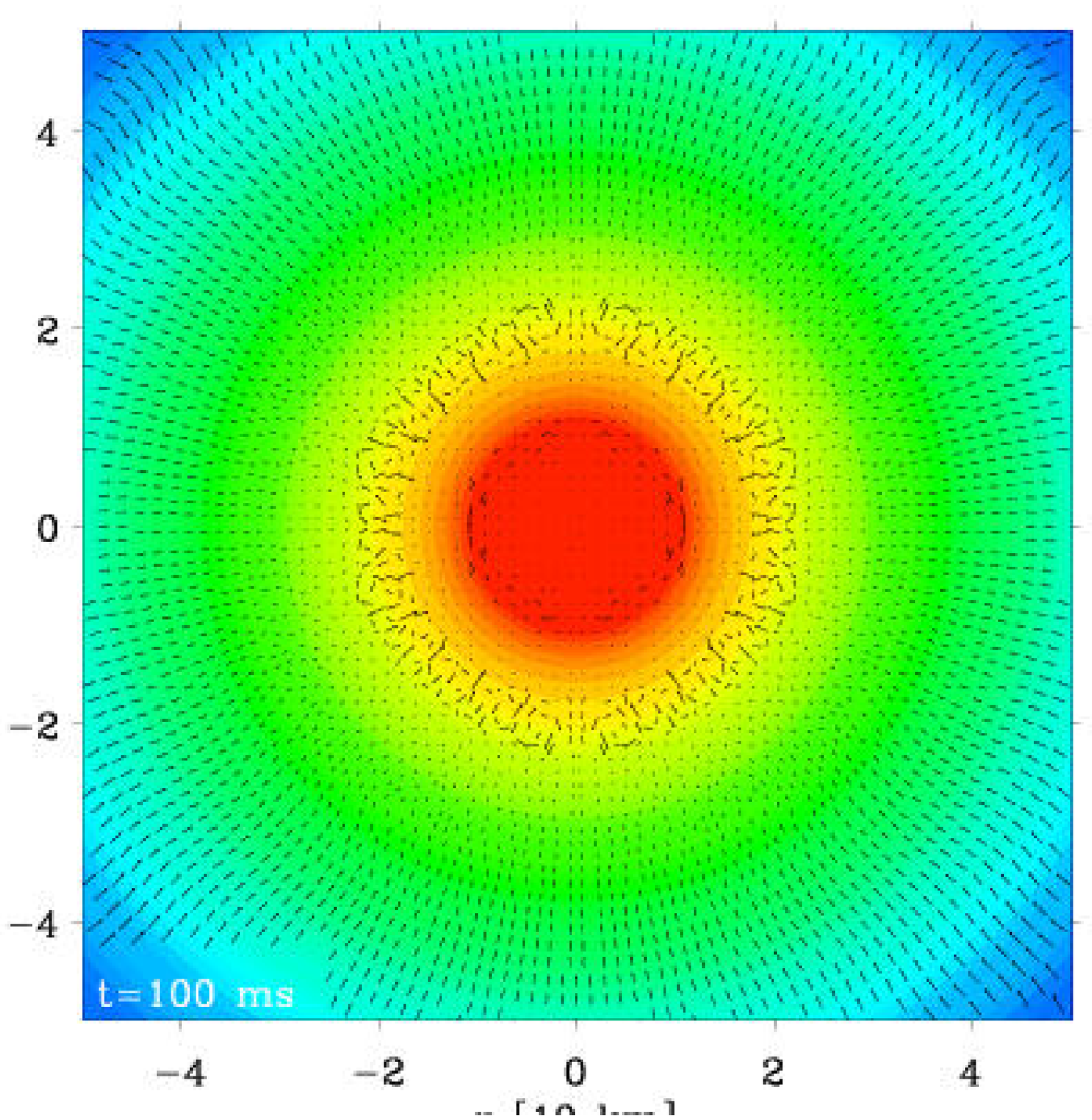}
\plottwo{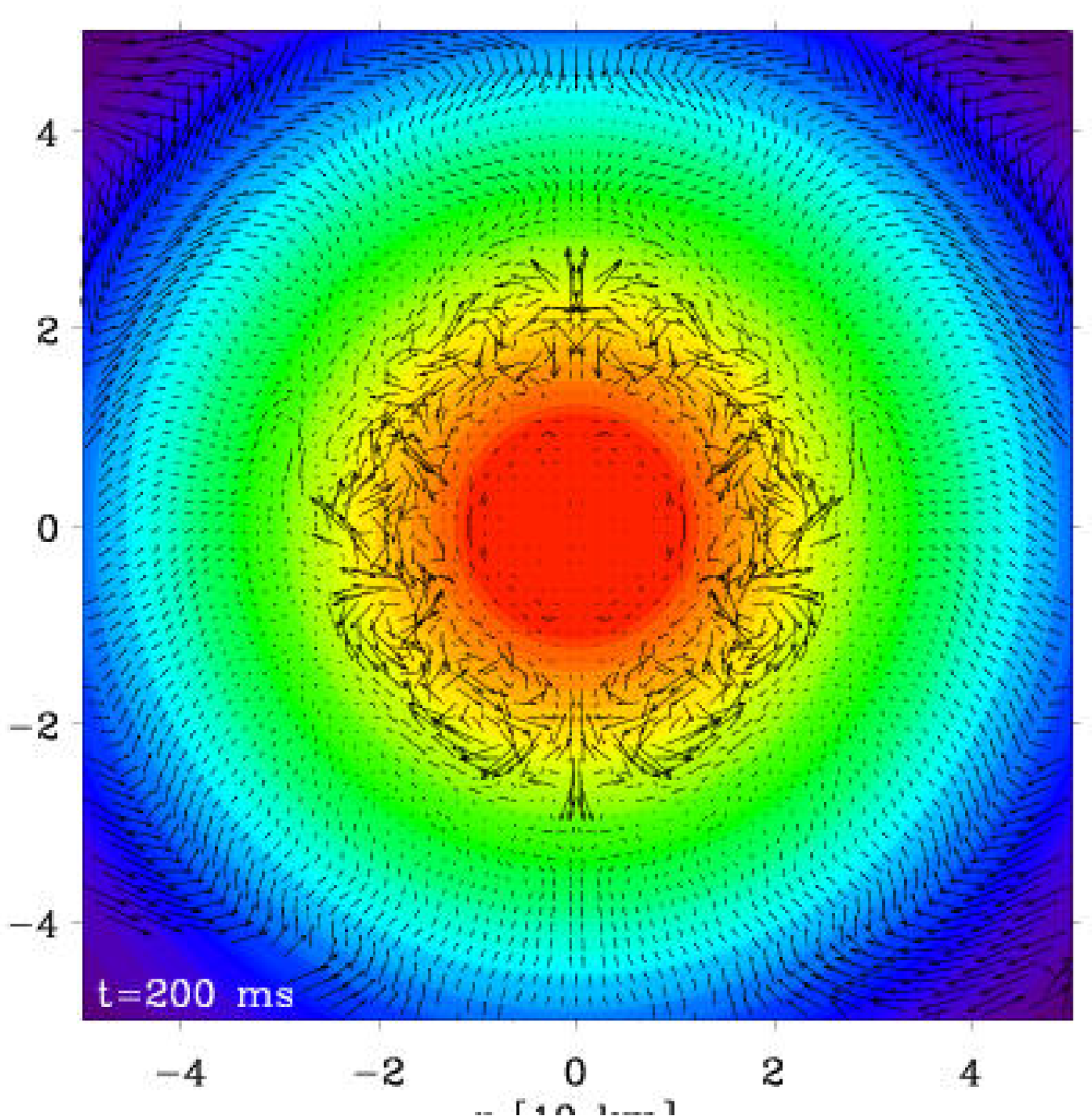}{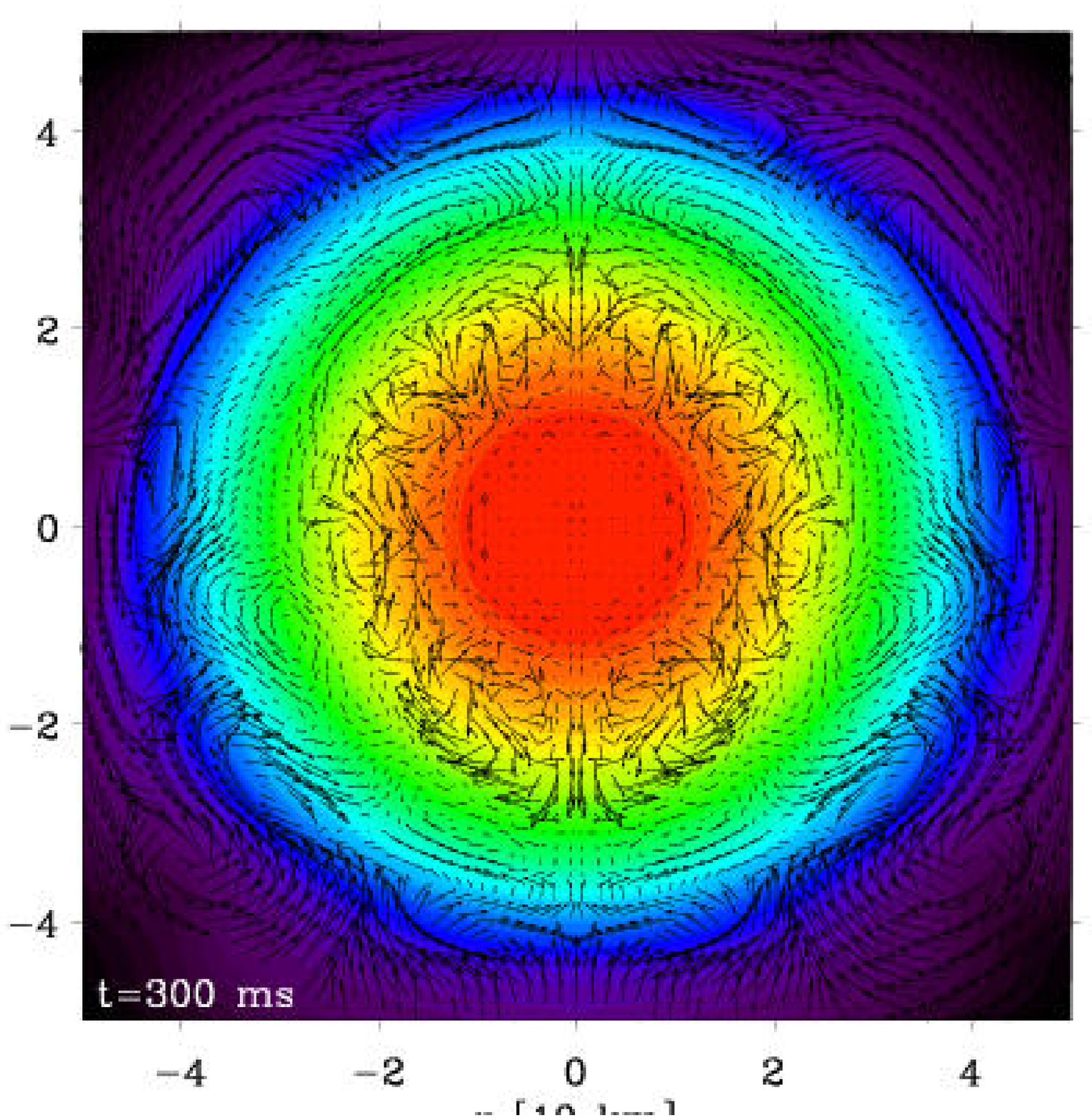}
\caption{
Same as Fig.~\ref{fig_entropy}, but for the mass density $\rho$.
(See text for discussion.)
}
\label{fig_density}
\end{figure*}

In this section, we present results from the baseline VULCAN/2D simulation, whose parameters and
characteristics were described in \S\ref{code}.
First, we present for the pre-bounce phase a montage 
of radial slices of the density (top-left), electron-fraction (top-right), 
temperature (middle-left), entropy (middle-right), radial-velocity (bottom-left), 
and Mach number (bottom right) in Fig.~\ref{fig_pre_cc}, using a solid line to represent 
the polar direction (90$^{\circ}$) and a dotted-line for the equatorial direction.
All curves overlap to within a line thickness, apart from the red curve, which corresponds
to the bounce-phase (within 1\,ms after bounce), showing the expected and trivial result that 
the collapse is indeed purely spherical.

\begin{figure*}
\plotone{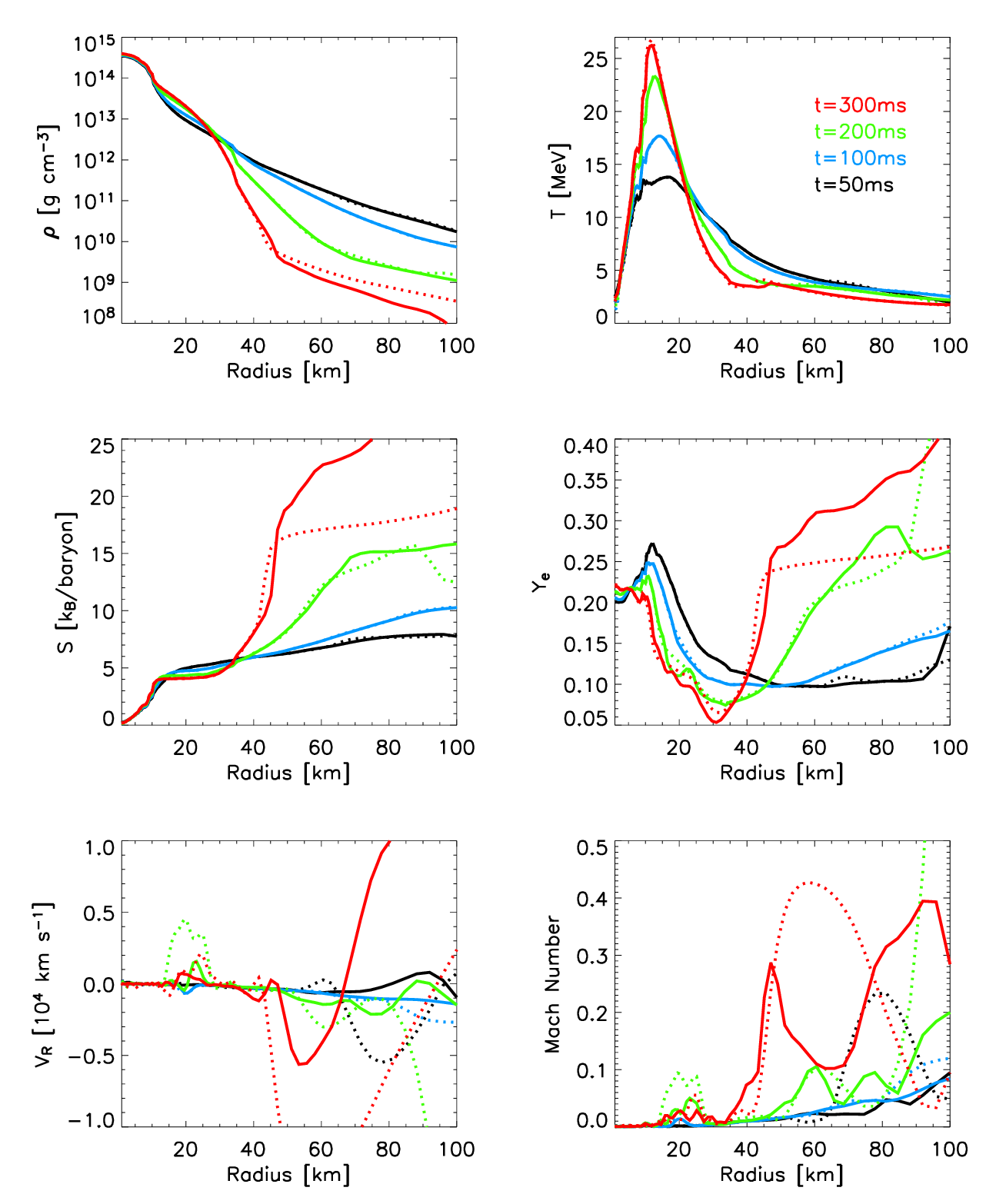}
\caption{Montage of radial cuts along the equatorial direction at 50 (black), 100 (blue), 200 (green),
and 300\,ms (red) past core bounce, echoing the properties displayed in 
Figs~\ref{fig_entropy}-\ref{fig_density} for the baseline model, for the density (top left), 
temperature (top right), entropy (middle left), $Y_{\rm e}$ (middle right), radial velocity 
(bottom left), and Mach number (bottom right).
}
\label{fig_4slices}
\end{figure*}

Now, let us describe the gross properties of the simulation, covering the first 300\,ms
past core bounce and focusing exclusively on the inner $\sim$50\,km.
In Figs.~\ref{fig_entropy}--\ref{fig_density}, we show stills of the entropy (Fig.~\ref{fig_entropy}), electron
fraction (Fig.~\ref{fig_ye}), and density (Fig.~\ref{fig_density}) at 50 (top left panel), 100 (top right),
200 (bottom left), and 300\,ms (bottom right) after core bounce.
We also provide, in Fig.~\ref{fig_4slices}, radial cuts of a sample of
quantities in the equatorial direction, to provide a clearer view of, for example, gradients.
Overall, the velocity magnitude is in excess of 4000\kms only beyond $\sim$50\,km, while it is systematically
below 2000\kms within the same radius.
At early times after bounce ($t=50$\,ms), the various plotted quantities are relatively similar throughout
the inner 50\,km. The material velocities are mostly radial, oriented inward, and very small, {\it i.e.}, 
do not exceed $\sim 1000$\,\kms. The corresponding Mach numbers throughout the PNS are 
subsonic, not reaching more than $\sim$10\% of the local sound speed.
This rather quiescent structure is an artefact of the early history of the young PNS before
vigorous dynamics ensues. The shock wave generated at core bounce, after the 
initial dramatic compression up to nuclear densities 
($\sim$3$\times$10$^{14}$ g\,cm$^{-3}$) of the inner progenitor regions, leaves a positive 
entropy gradient, reaching then its maximum of $\sim$6-7\,k$_{\rm B}$/baryon at $\sim$150\,km, 
just below the shock.
The electron fraction ($Y_{\rm e}$) shows a broad minimum between $\sim$30 and $\sim$90\,km, a result 
of the continuous deleptonization of the corresponding regions starting after the neutrino burst 
near core bounce.
Within the innermost radii ($\sim$10--20\,km), the very high densities ($\ge$ 10$^{12}$\,g\,cm$^{3}$)
ensure that the region is optically-thick to neutrinos, inhibiting their escape.

Turning to the next phase in our time series ($t=100$\,ms), we now clearly identify four zones within 
the inner 50\,km, ordered from innermost to outermost, which will
become increasingly distinct with time:
\begin{itemize}
\item Region A: This is the innermost region, within 10\,km, with an entropy of 
$\sim$1\,k$_{\rm B}$/baryon, a $Y_{\rm e}$ of 0.2--0.3, a density of $\sim$1-4$\times$10$^{14}$\,g\,cm$^{-3}$, 
essentially at rest with a near-zero Mach number (negligible vorticity and divergence).
This region has not appreciably changed in the elapsed 50\,ms, and will in fact not
do so for the entire evolution described here.
\item Region B: Between 10 and 30\,km, we see a region of higher entropy (2--5\,k$_{\rm B}$/baryon) with
positive-gradient and lower $Y_{\rm e}$ with negative gradient (from $Y_{\rm e}$ of $\sim$0.3 down to $\sim$0.1).
Despite generally low Mach number, this region exhibits significant motions with pronounced
vorticity, resulting from the unstable negative-gradient of the electron (lepton) fraction.
\item Region C: Between 30 and 50\,km is a region of outwardly-increasing entropy (5--8\,k$_{\rm B}$/baryon),
but with a flat and low electron fraction; this is the most deleptonized region in the entire
simulated object at this time. There, velocities are vanishingly small 
($\ll$\,1000\kms), as in Region A, although generally oriented radially inwards.
This is the cavity region where gravity waves are generated, most clearly at 200--300\,ms in our time series.
\item Region D: Above 50\,km, the entropy is still increasing outward, with values in excess
of 8\,k$_{\rm B}$/baryon, but now with an outwardly-increasing $Y_{\rm e}$ (from the minimum of 0.1 up to 0.2).
Velocities are much larger than those seen in Region B, although still corresponding to subsonic motions 
at early times. Negligible vorticity is generated at the interface between Regions C and D.
The radially infalling material is prevented from entering Region C and instead settles on its periphery.
\end{itemize}

\begin{figure*}
\plotone{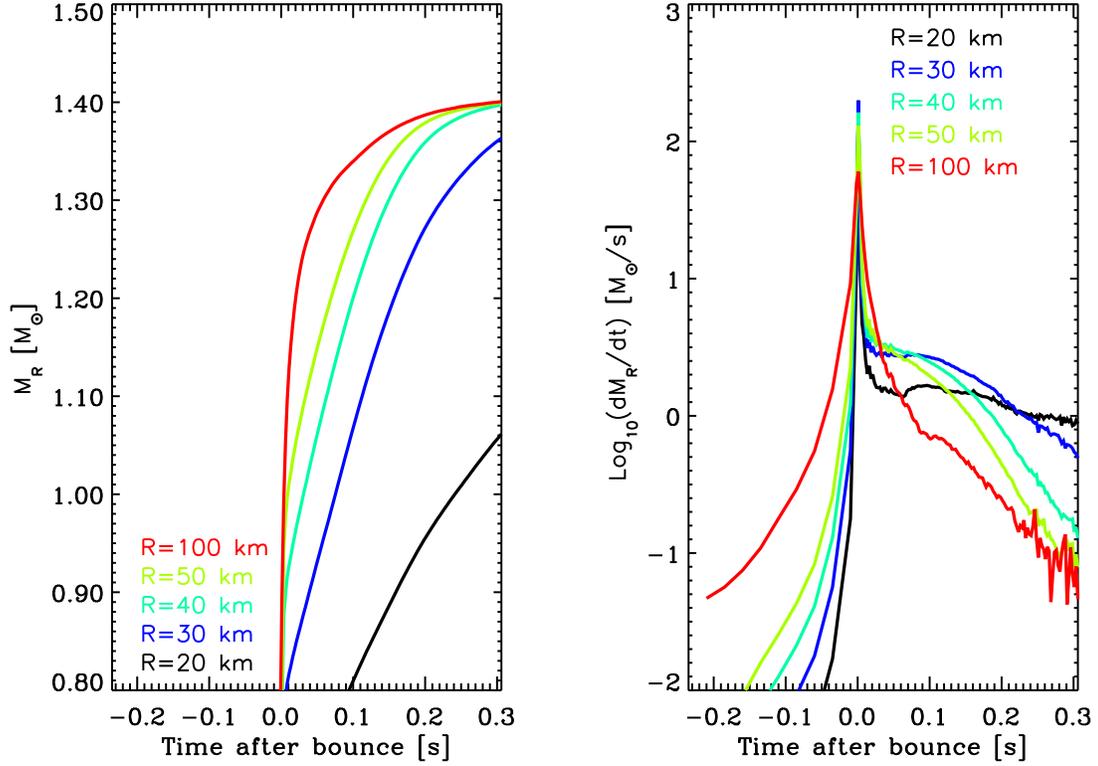}
\caption{{\it Left}: Time evolution after bounce of the interior mass spherical shells at 
selected radii: 20\,km (black), 30\,km (blue), 40\,km (red) and 50\,km (black). 
{\it Right}: Corresponding mass flow through the same set of radii.}
\label{fig_mdot_pns}
\end{figure*}

\begin{figure*}
\plotone{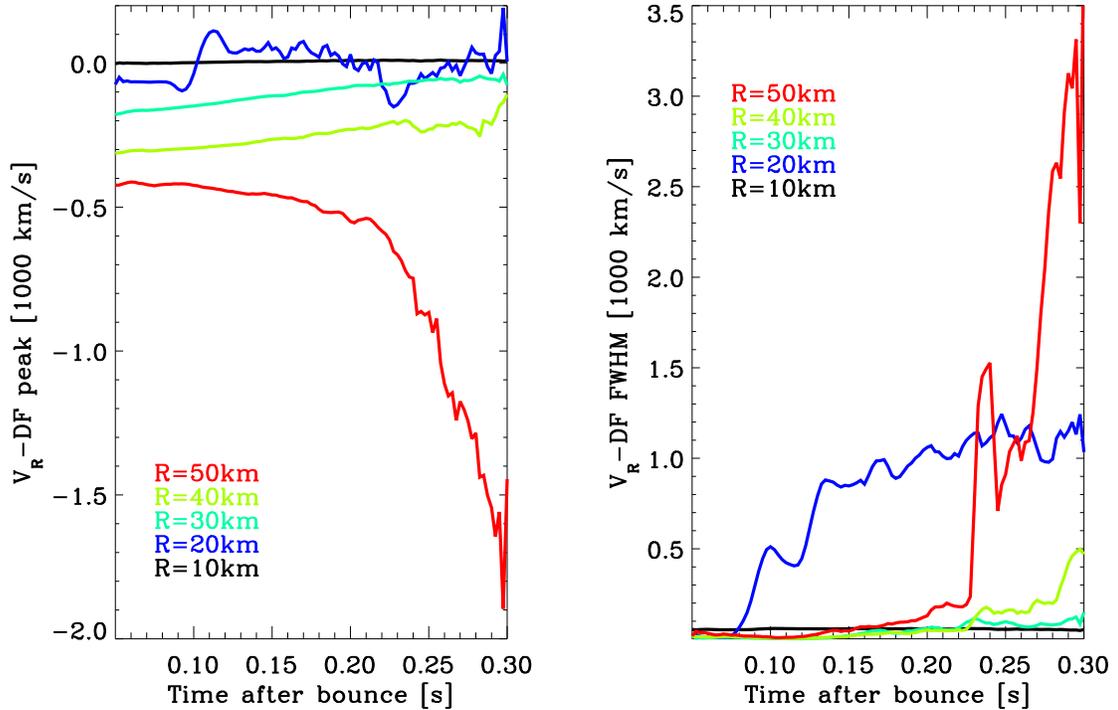}
\caption{
Time evolution, at selected radii, of the radial-velocity at peak and Full Width at Half 
Maximum (FWHM) of the radial-velocity distribution function. (See text for discussion.)
}
\label{fig_DF_Vr}
\end{figure*}

\begin{figure*}
\plotone{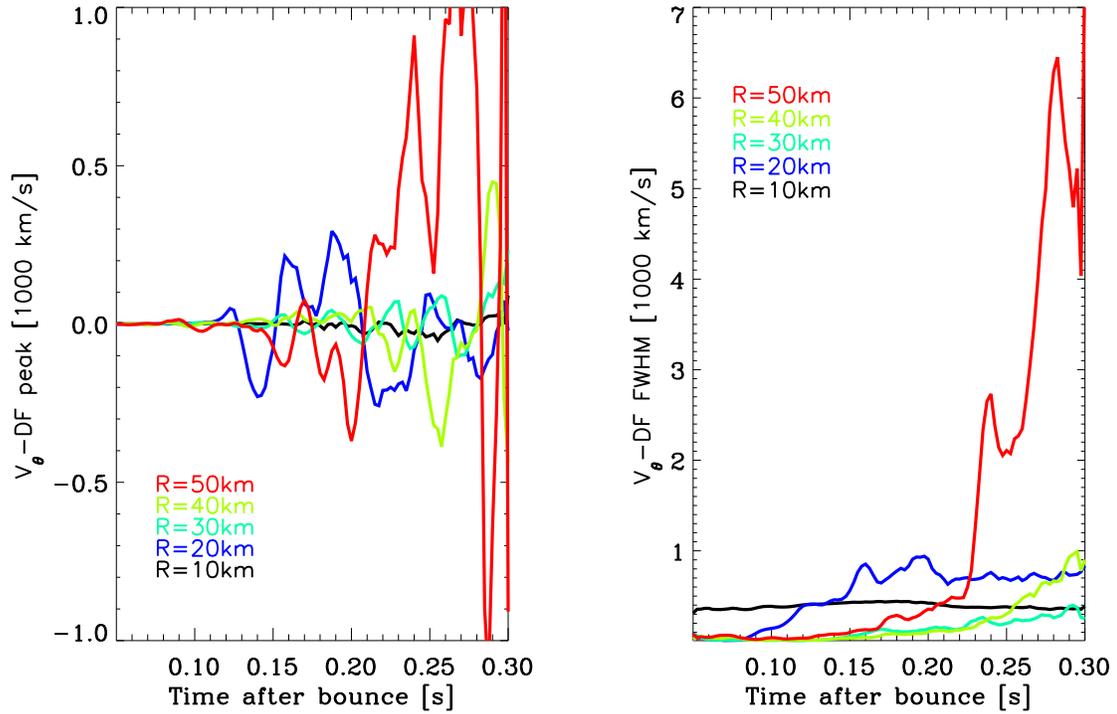}
\caption{
Same as Fig.~\ref{fig_DF_Vr}, but for the latitudinal velocity, $V_{\theta}$.
}
\label{fig_DF_Vt}
\end{figure*}

\begin{figure*}
\plotone{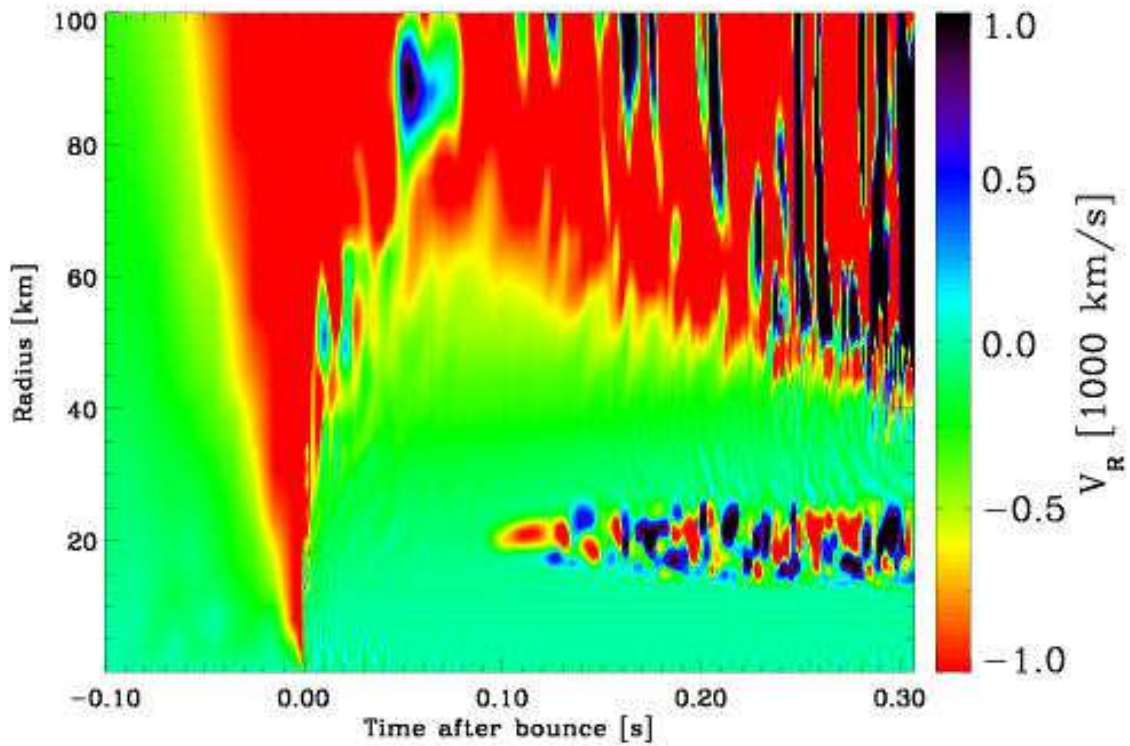}
\caption{
Color map of the radial velocity $V_R$ as a function of time after bounce and radius, along 
the equatorial direction.
The green regions denote relatively quiescent areas.  The inner region of the outer convective 
zone (Region D) is the predominately red zone; the horizontal band near $\sim$20 km is Region B, 
where isolated PNS convection obtains.  See Buras et al. (2005) for a similar plot and the text 
for details.
}
\label{fig_vr_rad}
\end{figure*}

\begin{figure*}
\plotone{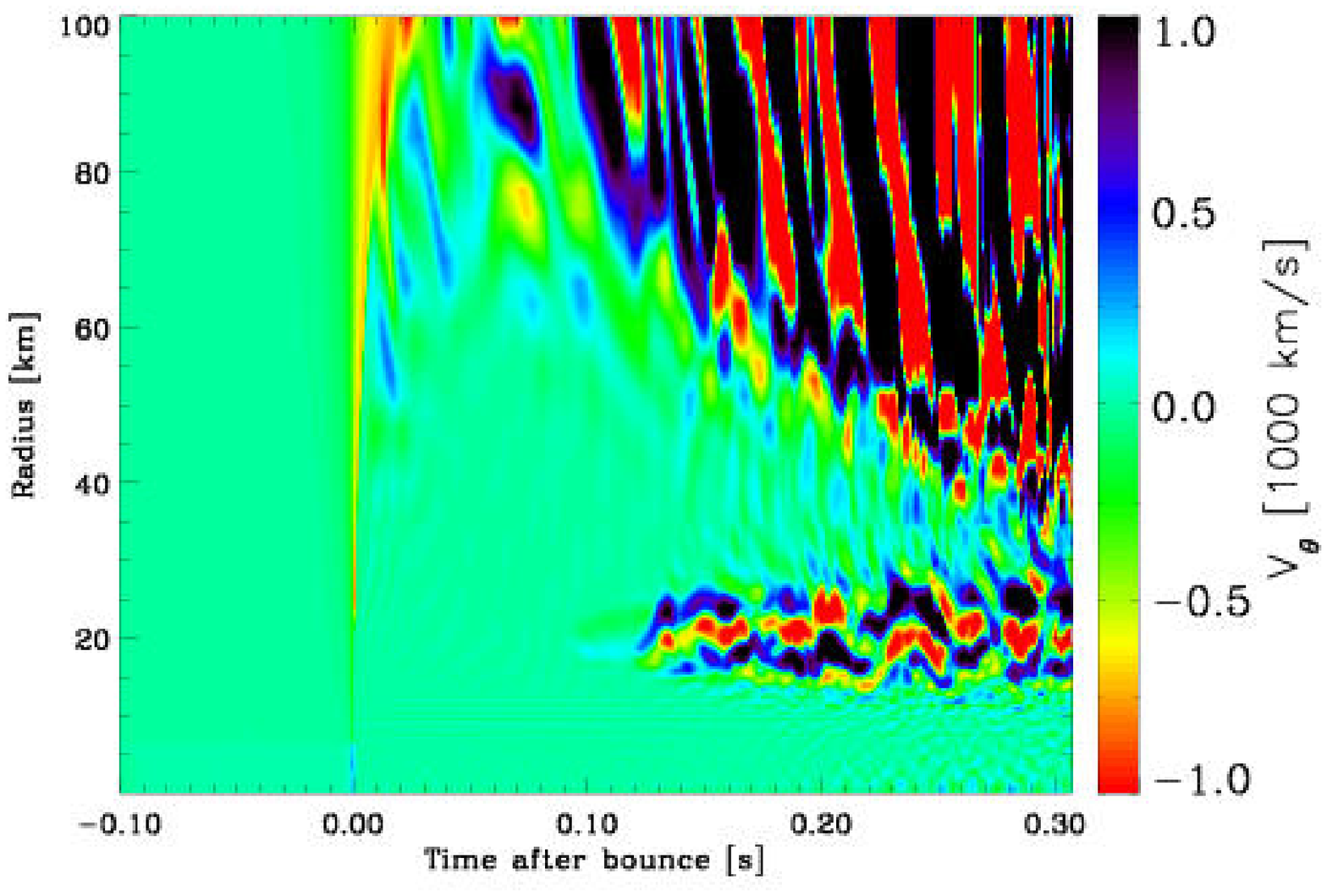}
\caption{
Same as Fig.~\ref{fig_vr_rad}, but for the latitudinal velocity ($V_{\theta}$).
Note the gravity waves excited between 30--40\,km, more visible in this image of the 
latitudinal velocity than in the previous figure for the radial velocity.
}
\label{fig_vt_rad}
\end{figure*}

As time progresses, these four regions persist, evolving only slowly for the entire $\sim$300\,ms after bounce. 
The electron fraction at the outer edge of Region A decreases.
The convective motions in low--$Y_{\rm e}$ Region B induce significant mixing of the high-$Y_{\rm e}$ interface
with Region A, smoothing the  $Y_{\rm e}$ peak at $\sim$10\,km.
Overall, Region A is the least changing region. 
In Region B, convective motions, although subsonic, are becoming more violent with time, reaching Mach numbers of
$\sim$0.1 at 200--300\,ms, associated with a complex flow velocity pattern.
In Region C, the trough in electron fraction becomes more pronounced, reaching down to 0.1 at 200\,ms,
and a record-low of 0.05 at 300\,ms. Just on its outer edge,
one sees sizable (a few$\times$100\kms) and nearly-exclusively 
latitudinal motions, persisting over large angular scales.
Region D has changed significantly, displaying low-density, 
large-scale structures with downward and upward velocities. 
These effectively couple remote regions, between the high-entropy, high--$Y_{\rm e}$ shocked region
and the low-entropy, low--$Y_{\rm e}$ Region C. Region D also stretches 
further in (down to $\sim$45\,km), at the expense of Region C which becomes more compressed.
This buffer region C seems to shelter the interior, which has changed more modestly than region D.

Figure~\ref{fig_4slices} shows radial cuts along the equator for the four time snapshots
(black: 50\,ms; blue: 100\,ms; green: 200\,ms; red: 300\,ms) shown
in Figs.~\ref{fig_entropy}-\ref{fig_density} for the density (upper left), temperature (upper right),
entropy (middle left), $Y_{\rm e}$ (middle right), radial velocity (bottom left), and Mach number (bottom right).
Notice how the different regions are clearly visible in the radial-velocity plot, showing regions
of significant upward/downward motion around 20\,km (Region B) and above $\sim$50\,km (Region D).
One can also clearly identify the $Y_{\rm e}$ trough, whose extent decreases from 30--90\,km at 50\,ms to
30--40\,km at 300\,ms.
The below-unity Mach number throughout the inner 50\,km also argues for little ram pressure associated with
convective motions in those inner regions.
Together with the nearly-zero radial velocities in Regions A and C at all times, 
this suggests that of these three regions mass motions
are confined to Region B.

One can identify a trend of systematic compression of Regions B--C--D,
following the infall of the progenitor mantle.
Indeed, despite the near-stationarity of the shock at 100--200\,km over the first 200--300\,ms, a large 
amount of mass flows inward through it, coming to rest at smaller radii.
In Fig.~\ref{fig_mdot_pns}, we display the evolution, up to 300\,ms past core bounce,
of the interior mass and mass flow through different radial shells within the PNS.
Note that mass inflow in this context has two components: 1) direct accretion of material from the 
infalling progenitor envelope and, 2) the compression of the PNS material as it cools and deleptonizes.
Hence, mass inflow through radial shells within the PNS would still be observed even in the absence of 
explicit accretion of material from the shock region. 
By 300\,ms past core bounce, the ``accretion rate'' has decreased from a maximum of $\sim$ 1-3\mdot at 
50\,ms down to values below 0.1\mdot and the interior mass at
30\,km has reached up to $\sim$1.36\,M$\sun$, i.e., 95\% of the progenitor core mass.
Interestingly, the mass flux at a radius of 20\,km is lower (higher) at early (late) times compared to that
in the above layers, and remains non-negligible even at 300\,ms.

An instructive way to characterize the properties of the fluid within the PNS is by means of
Distribution Functions (DFs), often employed for the description of the solar convection zone 
(Browning, Brun, \& Toomre 2004). Given a variable $x$ and a function $f(x)$, 
one can compute, for a range of values $y$ encompassing the extrema of $f(x)$, the new function
$$
g(f(x),y) \propto \exp \left[- \left(\frac{y-f(x)}{\sqrt{2}\sigma}\right)^2 \right],
$$
where $\sigma = \sqrt{<f^2(x)>_x-<f(x)>_x^2}$ and $<>_x$ is to be understood as an average over $x$.
We then construct the new function:
$$
h(y) = <g(f(x),y)>_x.
$$
Here, to highlight the key characteristics of various fluid quantities, we extract only the $y$ value 
$y_{\rm peak}$ at which $h(y)$ is maximum, {\it i.e.}, the most representative value $f(x)$ in our 
sample over all $x$ (akin to the mean),
and the typical scatter around that value, which we associate with the Full Width at Half Maximum (FWHM)
of the Gaussian-like distribution.

\begin{figure*}
\plotone{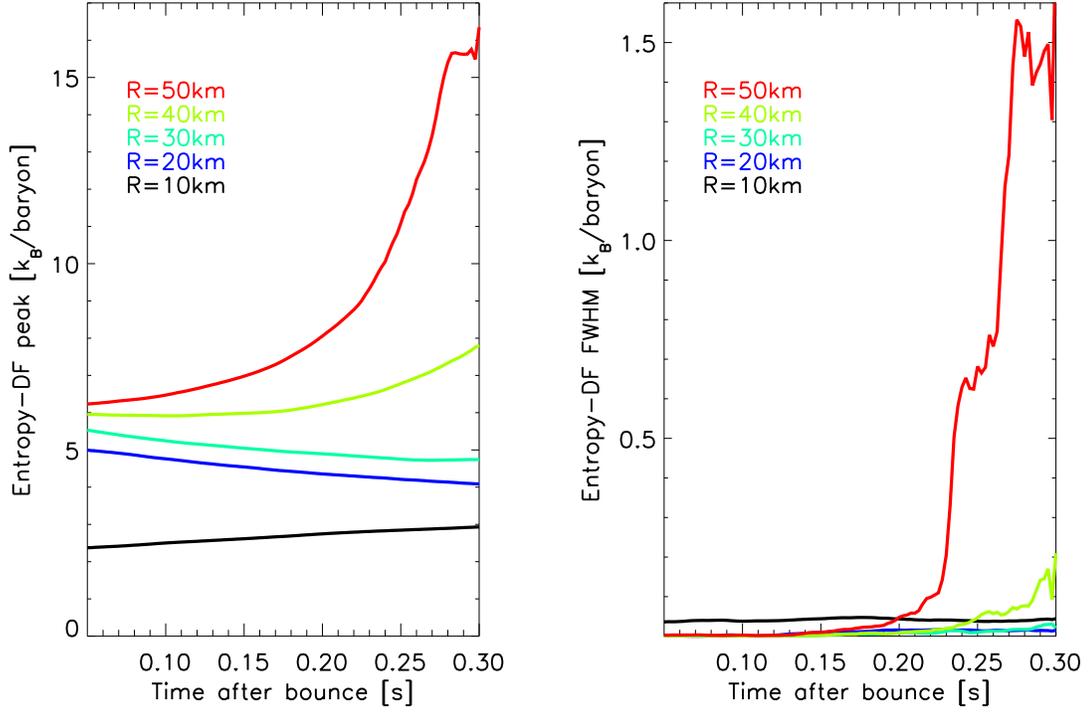}
\caption{
Time evolution after bounce, at selected radii, of the entropy at the peak (left) and FWHM
(right) of the entropy distribution function.
}
\label{fig_DF_S}
\end{figure*}

\begin{figure*}
\plotone{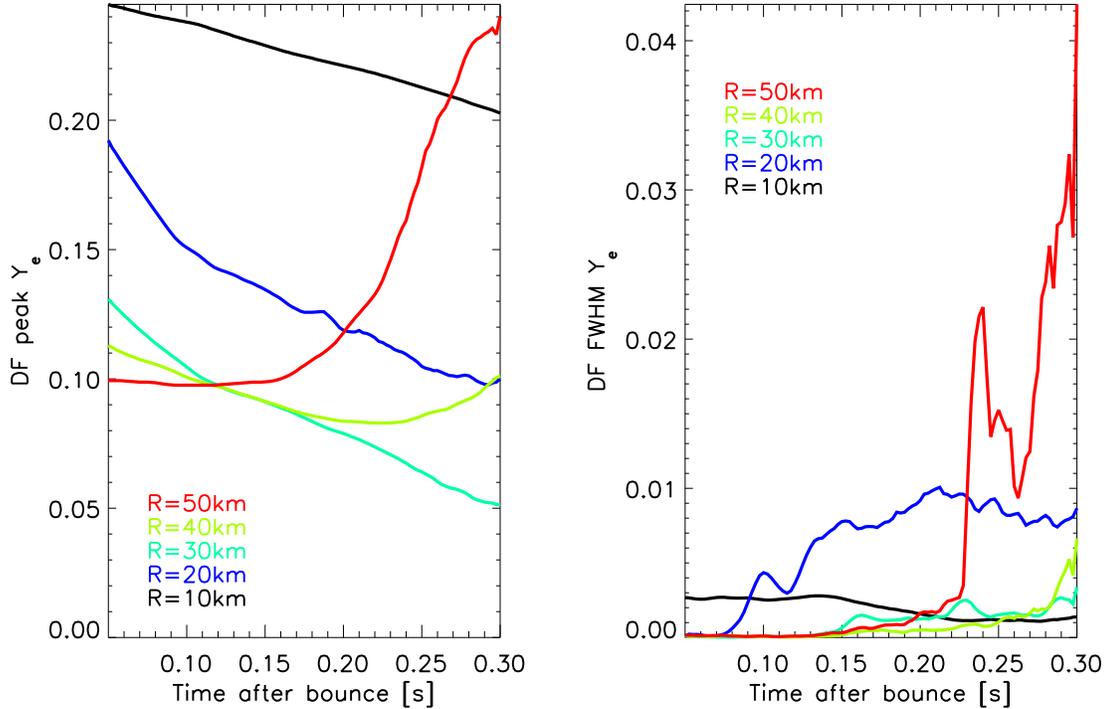}
\caption{Same as Fig.~\ref{fig_DF_S}, but for the electron fraction $Y_{\rm e}$.
}
\label{fig_DF_Ye}
\end{figure*}

In Figs.~\ref{fig_DF_Vr}-\ref{fig_DF_Vt} and \ref{fig_DF_S}-\ref{fig_DF_Ye}, 
we plot such peak and FWHM values at selected radii within the
PNS, each lying in one of the Regions A, B, C, or D, covering the times between 
50 and 300\,ms after core bounce.
Figure~\ref{fig_DF_Vr} shows the radial-velocity at the peak (left panel) 
and FWHM (right panel) of the radial-velocity distribution
function. The black, blue, turquoise, and green curves (corresponding to radii interior to 40\,km) are similar,
being close to zero for both the peak and the FWHM. In contrast, the red curve (corresponding
to a radius of 50\,km) shows a DF with a strongly negative peak radial velocity, even more so at later
times, while the FWHM follows the same evolution (but at positive values).
This is in conformity with the previous discussion.
Above $\sim$40\,km (Region D), convection underneath the
shocked region induces large-scale upward and downward motions, with velocities of a few 1000\kms, but 
negative on average, reflecting the continuous collapse of the progenitor mantle (Fig.~\ref{fig_mdot_pns}).
Below $\sim$40\,km, there is no sizable radial flow of material biased towards inflow, on average or at any time.
This region is indeed quite dynamically decoupled from 
the above regions during the first $\sim$300\,ms, in no obvious way 
influenced by the fierce convection taking place underneath the shocked region.
Turning to the distribution function of the latitudinal velocity
(Fig.~\ref{fig_DF_Vt}), we see a similar dichotomy between its peak and FWHM at radii below and above 40\,km.
At each radius, $V_{\theta}$ is of comparable magnitude to $V_R$, apart from the peak value which
remains close to zero even at larger radii (up to 40\,km).
This makes sense, since no body-force operates continuously pole-ward or equator-ward - the gravitational
acceleration acts mostly radially.
Radial- and latitudinal-velocity distribution functions are, therefore, strikingly similar at 10, 20, 30, and 40\,km,
throughout the first 300\,ms after core bounce, quite unlike the above layer, where the Mach number eventually
reaches close to unity between 50--100\,km (Fig.~\ref{fig_4slices}).
In these two figures, PNS convection is clearly visible at 20\,km, with small peak, 
but very sizable FWHM, values for the velocity distributions, highlighting the large scatter in
velocities at this height. Notice also the larger scatter of values for the lateral velocity
at 10\,km in Fig.~\ref{fig_DF_Vt}, related to the presence of the horns and the transition
radius at this height.

In Figs.~\ref{fig_vr_rad}--\ref{fig_vt_rad}, we complement the previous 
two figures by showing the temporal evolution of the radial and latitudinal 
velocities, using a sampling of one millisecond, along the equatorial direction and over the inner 100\,km.
To enhance the contrast in the displayed properties, we cover the entire evolution computed with VULCAN/2D,
from the start at 240\,ms prior to, until 300\,ms past core bounce.
Note the sudden rise at $\sim$0.03\,s prior to bounce, stretching down to radii
of $\sim$2-3\,km, before the core reaches nuclear densities and bounces.
The shock wave moves out to $\sim$150\,km (outside of the range shown), where it stalls.
In the 50-100\,km region along the equator, we observe mostly downward motions, which reflect the systematic
infall of the progenitor envelope, but also the fact that upward motions (whose presence is 
inferred from the distribution function of the radial velocity) occur preferentially at non-zero latitudes.
The minimum radius reached by these downward plumes decreases with time, from $\sim$70\,km at 100\,ms down
to $\sim$40\,km at 300\,ms past core bounce. 
Note that these red and blue ``stripes'' are slanted systematically towards smaller heights for increasing
time, giving corresponding velocities $\Delta r / \Delta t \sim -50\,{\rm km}/10\,{\rm ms} \sim -5000$\kms, in agreement
with values plotted in Fig.~\ref{fig_4slices}.
The region of small radial infall above 30\,km and extending to 60--35\,km (from 50 to 300\,ms past core bounce)
is associated with the trough in the $Y_{\rm e}$ profile (Fig.~\ref{fig_4slices}), narrowing significantly as the envelope
accumulates in the interior (Fig.~\ref{fig_mdot_pns}).
The region of alternating upward and downward motions around 20\,km persists there at all times followed, confirming
the general trend seen in Figs.~\ref{fig_DF_Vr}-\ref{fig_DF_Vt}.
The inner 10\,km (Region A) does not show any appreciable motions at any time, even with this very fine time sampling.
The latitudinal velocity displays a similar pattern (Fig.~\ref{fig_vt_rad}) to that of the radial velocity,
showing time-dependent patterns in the corresponding regions.
However, we see clearly a distinctive pattern after 100\,ms past bounce and above $\sim$50\,km, recurring periodically
every $\sim$15\,ms.
This timescale is comparable to the convective overturn time for downward/upward plumes moving back and forth between
the top of Region C at $\sim$50\,km and the shock region at 150\,km, with typical velocities of 5000\kms, {\it i.e.}
$\tau \sim 100$km/5000\kms $\sim$ 20\,ms.
In Region C, at the interface between the two convective zones B and D, the latitudinal velocity $V_{\theta}$ 
has a larger amplitude and shows more time-dependence than the radial velocity $V_R$ in the corresponding region.
Interestingly, the periodicity of the patterns discernable in the $V_{\theta}$ field in Region C 
seems to be tuned to that in the convective Region D above, visually striking when one extends the 
slanted red and blue ``stripes'' from the convective Region D downwards to radii of $\sim$30\,km.   
This represents an alternative, albeit heuristic, demonstration of the potential excitation of
gravity waves in Region C by the convection occurring above (see \S\ref{grav_waves}).
What we depict in Figs.~\ref{fig_vr_rad}--\ref{fig_vt_rad} is also seen in Fig.~29 of 
Buras et al. (2005), where, for their model s15Gio\_32.b that switches from 2D to spherical-symmetry 
in the inner 2\,km, PNS convection obtains $\sim$50\,ms after bounce and between 10--20\,km.
The similarity between the results in these two figures indicates that as far as PNS convection 
is concerned and besides differences in approaches, VULCAN/2D and MUDBATH compare well.
Differences in the time of onset of PNS convection may be traceable solely to differences
in the initial seed perturbations, which are currently unknown.
In MUDBATH, the initial seed perturbations are larger than those inherently 
present in our baseline run, leading to an earlier onset by $\sim$50\,ms of the PNS convection
simulated by Buras et al. (2005).

We show the distribution function for the entropy in Fig.~\ref{fig_DF_S}. 
Again, we see both in the entropy at the peak and the FWHM the dichotomy 
between the inner 30\,km with low values, and the layers above, with much larger 
values for both.
All radii within the PNS start, shortly after core bounce (here, the initial 
time is 50\,ms), with similar values, around 5-7\,k$_{\rm B}$/baryon.
Below 20\,km, convective motions homogenize the entropy, giving the
very low scatter, while the relative isolation of these regions from the convection and net neutrino 
energy deposition above maintains the peak value low. Outer regions (above 30\,km) are 
considerably mixed with accreting material, enhancing the entropy 
considerably, up to values of 20-30\,k$_{\rm B}$/baryon.

To conclude this descriptive section, we show in Fig.~\ref{fig_DF_Ye} 
the distribution function for the electron fraction. The dichotomy reported 
above between different regions is present here. Above $\sim$30\,km, the $Y_{\rm e}$ increases
with time as fresh material accretes from larger radii, while below this limit, the absent or
modest accretion cannot compensate for the rapid electron capture and neutrino losses.
Indeed, the minimum at 30\,km and 300\,ms corresponds roughly with the position of the 
neutrinosphere(s) at late times, which is then mostly independent of neutrino energy (\S\ref{pns}).

\section{Protoneutron Star Convection, Doubly-Diffusive Instabilities, and Gravity Waves}
\label{pns}

   In this section, we connect the results obtained for our baseline model to a number of potential modes and
instabilities that can arise within or just above the PNS (here again, we focus on the innermost 50--100\,km),
all related to the radial distribution of entropy and lepton number (or electron) fraction.
Instead of a stable combination of negative entropy gradient and positive $Y_{\rm e}$ gradient in
the PNS, the shock generated at core bounce leaves
a positive entropy gradient in its wake, while the concomitant deleptonization due to neutrino losses
at and above the neutrinosphere establishes a negative Y$_e$ gradient.
This configuration is unstable according to the Ledoux criterion and sets the background for our 
present discussion of PNS convection and motions.

\subsection{Protoneutron Star Convection}

\begin{figure*} 
\plotone{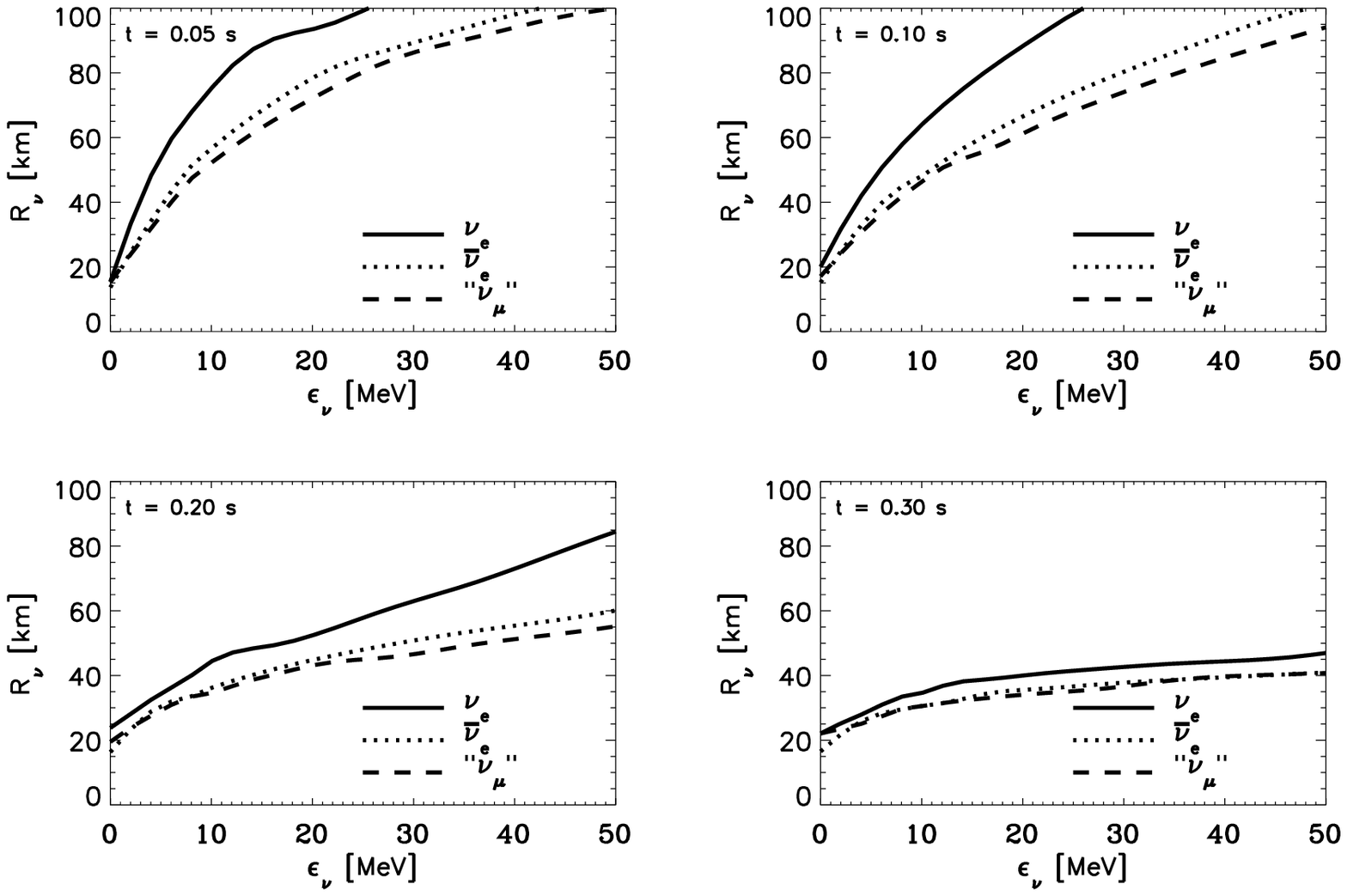}
\caption{Neutrino energy  dependence of neutrinosphere radii at four selected times past core bounce
($t=50$\,ms: top left; $t=100$\,ms: top right; $t=200$\,ms: bottom left; $t=300$\,ms: bottom right)
for the three neutrino ``flavors'' (${\nu_{\rm e}}$, solid line; ${\bar{\nu}}_{\rm e}$, 
dotted line; ``${\nu_{\mu}}$'', dashed line) treated in our 16-energy-group baseline simulation.
}
\label{fig_nu_sphere}
\end{figure*}

\begin{figure*}[htbp!]
\plotone{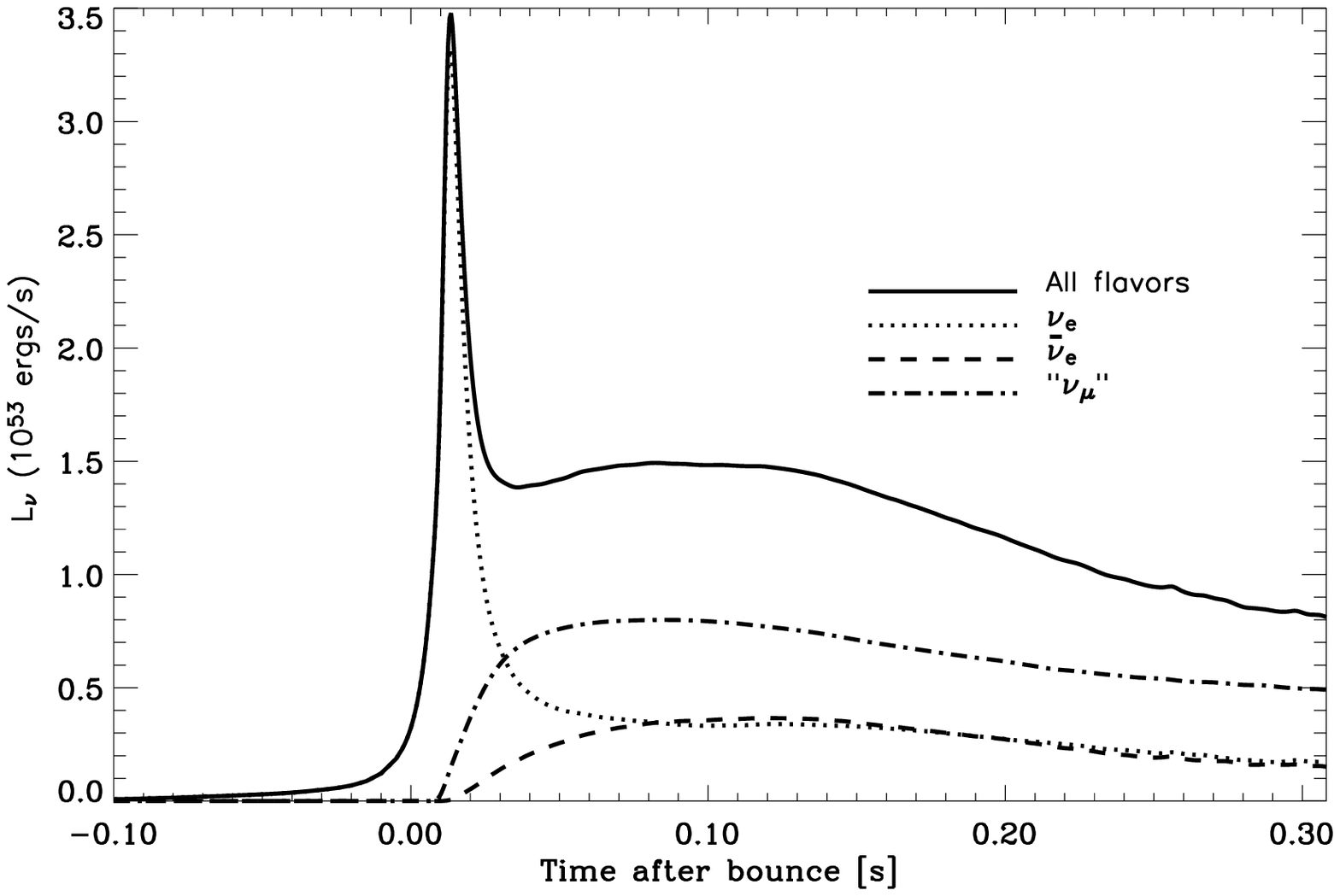}
\caption{
Time evolution of the neutrino luminosity, free-streaming through the outer grid radius 
at 3800\,km, for the three neutrino ``flavors'' (${\nu_{\rm e}}$, dotted line; 
${\bar{\nu}}_{\rm e}$, dashed line; ``${\nu_{\mu}}$'', dash-dotted line), as well as 
the sum of all three contributions (solid line), for our 16-energy-group baseline model. 
The time sampling is every 5\,ms until $t=-6$\,ms, and every 0.5\,ms for the remaining of 
the simulation, where we have defined $t=0$ as the time of hydrodynamical bounce. 
}
\label{fig_lumin}
\end{figure*}

In the preceding sections, we have identified two regions where sizable velocities persist over hundreds of
milliseconds, associated with the intermediate Region B (covering the range 10--20\,km) and the outer Region D
considered here (above $\sim$50\,km).
The latter is the region of advection-modified, neutrino-driven turbulent convection bounded by the shock wave.
While this region does not directly participate in the PNS convection, it does influence the interface
layer (Region C) and excites the gravity waves seen there (\S\ref{grav_waves}).

At intermediate radii (10--20\,km, Region B), we have identified a region 
of strengthening velocity and vorticity,
with little net radial velocity ($\le$\,100\kms) at a given radius when averaged over all directions. 
In other words, significant motions are seen, but confined within a small
region of modest radial extent of 10-20\,km at most.
As clearly shown in Fig.~\ref{fig_4slices}, this region has a rather flat entropy gradient, which cannot stabilize
the steep negative--$Y_{\rm e}$ gradient against this inner convection.
This configuration is unstable according to the Ledoux criterion
and had been invoked as a likely site of convection (Epstein 1979; Lattimer \& Mazurek 1981; Burrows 1987).
It has been argued that such convection,
could lead to a sizable advection of neutrino energy upwards, into regions where neutrinos are decoupled from
the matter ({\it i.e.}, with both low absorptive and scattering opacities), thereby making promptly available  
energy which would otherwise have diffused out over a much longer time.

The relevance of any advected flux in this context rests on whether the neutrino energy is advected
from the diffusive to the free-streaming regions, {\it i.e.}, whether some material is indeed dredged from 
the dense and neutrino-opaque core out to and above the neutrinosphere(s).
In Fig.~\ref{fig_nu_sphere}, we show the energy-dependent neutrinospheric radii 
$R_{\nu}(\epsilon_{\nu})$ at four 
selected times past core bounce (50, 100, 200, and 300\,ms) for the three neutrino ``flavors'' 
(${\nu_{\rm e}}$, solid line; $\bar{\nu}_{\rm e}$,
dotted line; ``${\nu_{\mu}}$'', dashed line), with $R_{\nu}(\epsilon_{\nu})$ defined by the relation,
$$ \tau(R_{\nu}(\epsilon_{\nu})) =\int_{R_{\nu}(\epsilon_{\nu})}^\infty 
                                   \kappa_{\nu}(\rho,T,Y_{\rm e}) \rho(R') dR' = 2/3\,,$$
where the integration is carried out along a radial ray. 
At 15\,MeV, near where the $\nu_{\rm e}$ 
and ${\bar{\nu}}_{\rm e}$ energy distributions at infinity peak, matter and neutrinos decouple at a radius
of $\sim$80\,km at $t=50$\,ms,  decreasing at later times to $\sim$30\,km.
Note that the neutrinospheric radius becomes less and less dependent on 
the neutrino energy as time proceeds, which results from the steepening
density gradient with increasing time (compare the black and red curves 
in the top left panel of Fig.~\ref{fig_4slices}).
This lower-limit on the neutrinospheric radius of 30\,km is to be compared 
with the 10--20\,km radii where PNS convection
obtains. The ``saddle'' region of very low $Y_{\rm e}$ at $\sim$30\,km, 
which harbors a very modest radial velocity at all times and, thus,
hangs steady, does not let any material penetrate through. 

Figure \ref{fig_lumin} depicts the neutrino luminosities until
300\,ms after bounce, showing, after the initial burst of the $\nu_e$ luminosity,  
the rise and decrease of the $\bar{\nu}_e$ and ``$\nu_{\mu}$'' luminosities
between 50 and 200\,ms after core bounce.
Compared to 1D simulations with SESAME for the same progenitor (Thompson et al. 2003), 
the $\bar{\nu}_e$ and ``$\nu_{\mu}$'' luminosities during this interval are larger 
by 15\% and 30\%, respectively.
Moreover, in the alternate model using a transition radius at 30\,km, we find 
enhancements of the same magnitudes and for the same neutrino species, but occurring 
$\sim$50\,ms earlier. This reflects the influence of the additional seeds introduced
by the horns located right in the region where PNS convection obtains.
We can thus conclude that the 15\% and 30\% enhancements in the $\bar{\nu}_e$ and ``$\nu_{\mu}$'' 
luminosities between 50 and 200\,ms after core bounce observed in our baseline model
are directly caused by the PNS convection.

There is evidence in the literature for enhancements of similar magnitude in post-maximum neutrino
luminosity profiles, persisting over $\sim$100\,ms, but associated with large modulations 
of the mass accretion rate, dominating the weaker effects of PNS convection.
In their Fig.~39, Buras
et al. (2005) show the presence of a $\sim$150\,ms wide bump in the luminosity of all
three neutrino flavors in their 2D run (the run with velocity terms omitted in the
transport equations). Since this model exploded after two hundred milliseconds,
the decrease in the luminosity (truncation of the bump) results then from the reversal
of accretion. Similar bumps are seen in the 1D code VERTEX 
(Rampp \& Janka 2002), during the same phase and for the same duration, as shown in the 
comparative study of Liebendorfer et al. (2005), but here again, associated
with a decrease in the accretion rate. In contrast, in their Fig.~10b, the ``$\nu_{\mu}$''
luminosity predicted by AGILE/BOLTZTRAN (Mezzacappa \& Bruenn 1993; Mezzacappa \& Messer 1999;
Liebendorfer et al. 2002,2004) does not show the post-maximum bump, although the 
electron and anti-electron neutrinos luminosities do show an excess similar to what VERTEX 
predicts. These codes are 1D and thus demonstrate that such small-magnitude 
bumps in neutrino luminosity may, in certain circumstances, stem from accretion.
Our study shows that in this case, the enhancement, of the $\bar{\nu}_e$  and 
``$\nu_{\mu}$'' luminosities, albeit modest, is due to PNS convection.

From the above discussion, we find that PNS convection causes the $\sim$200\,ms-long 10--30\% 
enhancement in the post-maximum $\bar{\nu}_e$ and ``$\nu_{\mu}$'' neutrino luminosities in our
baseline model, and thus we conclude that there is no sizable or 
long-lasting convective boost to the $\nu_e$ and $\bar{\nu}_e$ neutrino luminosities of
relevance to the neutrino-driven supernova model, and that what boost there may be
evaporates within the first $\sim$200\,ms of bounce.

\subsection{Doubly-diffusive instabilities}

When the medium is stable under the Ledoux criterion,
Bruenn, Raley, \& Mezzacappa (2005) argue for the potential presence
in the PNS of doubly-diffusive instabilities associated with gradients in electron fraction and entropy.
Whether doubly-diffusive instabilities occur is contingent upon
the diffusion timescales of composition and heat, mediated in the PNS by neutrinos.
Mayle \& Wilson (1988) suggested that so-called ``neutron fingers" existed in the PNS,
resulting from the fast transport of heat and the slow-equilibration transport of leptons.
This proposition was rejected by Bruenn \& Dineva (1996), who argued that these rates are in fact
reversed for two reasons. Energy transport by neutrinos, optimal in principle for higher-energy
neutrinos, is less impeded by material opacity at lower neutrino energy. Moreover, because lepton number is the
same for electron/anti-electron neutrinos {\it irrespective} of their energy, lepton transport
is faster than that of heat. This holds despite the contribution of the other neutrino types (which suffer
lower absorption rates) to heat transport.  

Despite this important, but subtle, difference in diffusion timescales for thermal and lepton
diffusion mediated by neutrinos, the presence of convection within the PNS, which operates
on much smaller timescales ({\it i.e.}, $\sim$\,1\,ms compared to $\sim$\,1\,s), outweighs these in importance.
In the PNS (a region with high neutrino absorption/scattering cross sections),
the presence of convection operating on timescales of the order of a few milliseconds seems to dominate
any doubly-diffusive instability associated with the transport of heat and leptons by neutrinos.
Furthermore, and importantly, we do not see overturning motions in regions not unstable to Ledoux convection.
Hence, we do not, during these simulations, discern the presence of doubly-diffusive instabilities at all.
This finding mitigates against the suggestion that doubly-diffusive instabilities in the first
300-500 milliseconds after bounce might perceptively enhance the neutrino luminosities 
that are the primary agents in the neutrino-driven supernova scenario.

\subsection{Gravity waves}
\label{grav_waves}

\begin{figure*}
\plotone{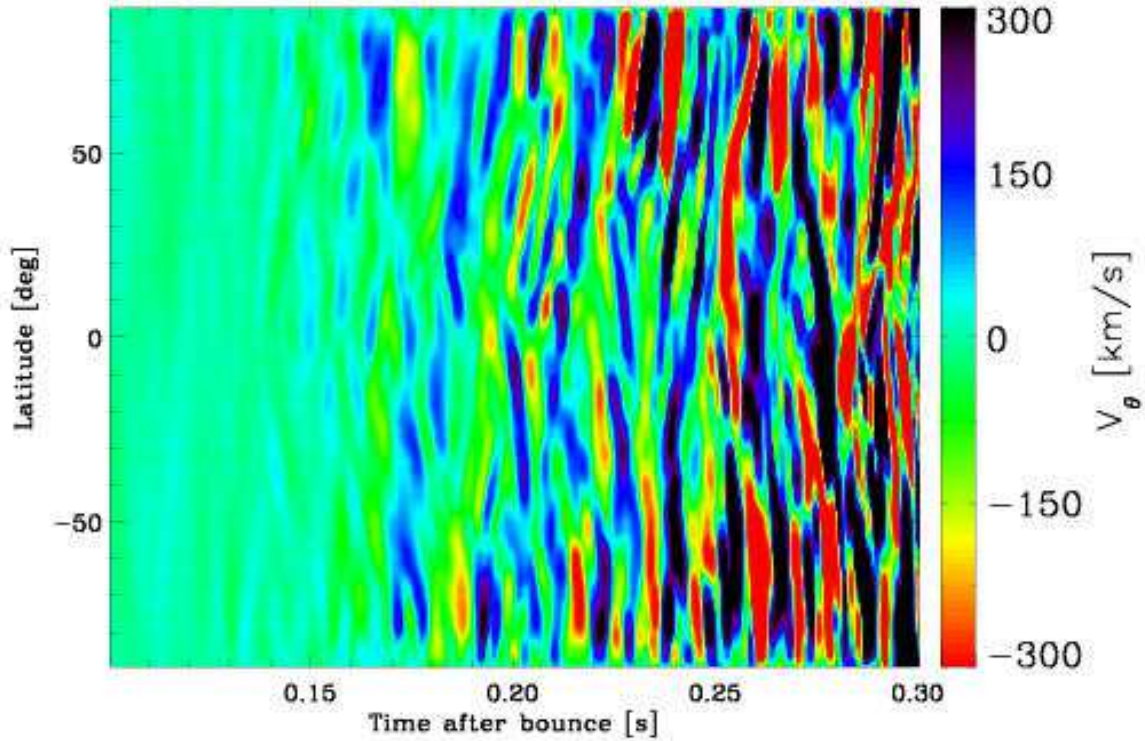}
\caption{
Color map of the latitudinal velocity ($V_{\theta}$) as a function of time after bounce 
and latitude, at a radius of 35\,km.
We choose a time range of 200\,ms, starting 100\,ms after core bounce, to show
the appearance, strengthening, and evolution to higher frequencies of gravity waves.
Also, due to accretion/compression, the corresponding region recedes to 
greater depths with increasing time.
}
\label{fig_vt_ang_gmodes}
\end{figure*}

We have described above the presence of a region (C) with little or no inflow up to 100--300\,ms 
past core bounce, which closely matches the location of minimum $Y_{\rm e}$ values in our simulations, 
whatever the time snapshot considered.
At such times after core bounce, we clearly identify the presence of waves at the corresponding height of
$\sim$30\,km, which also corresponds to the surface of the PNS where the density gradient steepens
(see top-left panel of Fig.~\ref{fig_4slices}).
As shown in Fig.~\ref{fig_nu_sphere}, as time progresses, this
steepening of the density profile causes the (deleptonizing) neutrinosphere to move inwards, converging to a
height of $\sim$30-40\,km, weakly dependent on the neutrino energy.

\begin{figure}
\plotone{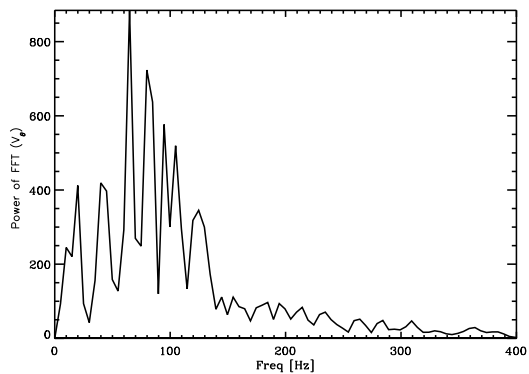}
\caption{Temporal spectrum of the latitudinal velocity ($V_{\theta}$) at a radius of 35\,km, averaged over all
directions, built from a sample of 401 frames equally spaced over the range 0.1--0.3\,s past core bounce.
}
\label{fig_power_fft_vt}
\end{figure}

To diagnose the nature and character of such waves, we show in Fig.~\ref{fig_vt_ang_gmodes} the fluctuation of
the latitudinal velocity (subtracted from its angular average), as a function of time and latitude, and at a
radius of 35\,km.
The time axis has been deliberately reduced to ensure that the region under scrutiny does not move inwards
appreciably during the sequence (Fig.~\ref{fig_mdot_pns}).
Taking a slice along the equator, we see a pattern with peaks and valleys repeated every $\sim$10--20\,ms,
with a smaller period at later times, likely resulting from the more violent development of the
convection underneath the shock.
We provide in Fig.~\ref{fig_power_fft_vt} the angular-averaged temporal power spectrum of the
latitudinal velocity (minus the mean in each direction).
Besides the peak frequency at 65\,Hz ($P\sim$15\,ms), we also observe its second harmonic
at $\sim$130\,Hz ($P\sim$7.5\,ms) together with intermediate frequencies at $\sim$70\,Hz ($P\sim$14\,ms), 
$\sim$100\,Hz ($P\sim$10\,ms), $\sim$110\,Hz ($P\sim$9\,ms).
The low frequencies at $\sim$20\,Hz ($P\sim$50\,ms), $\sim$40\,Hz ($P\sim$25\,ms) may stem from
the longer-term variation of the latitudinal velocity variations.

\begin{figure}
\plotone{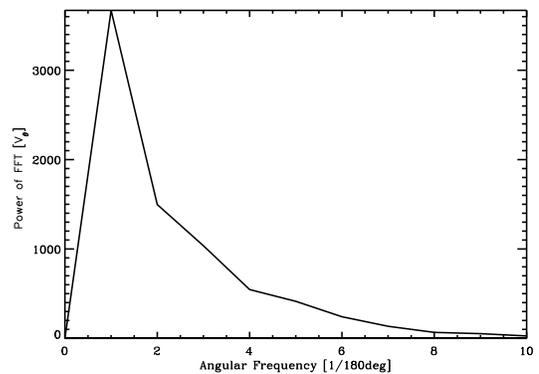}
\caption{
Angular spectrum of the Fourier Transform of the latitudinal velocity ($V_{\theta}$) 
at a radius of 35\,km, averaged over all times, built from a sample of 100 frames covering, 
with equal spacing, the angular extent of the grid.
}
\label{fig_power_fft_vt_lat}
\end{figure}

\begin{figure}
\plotone{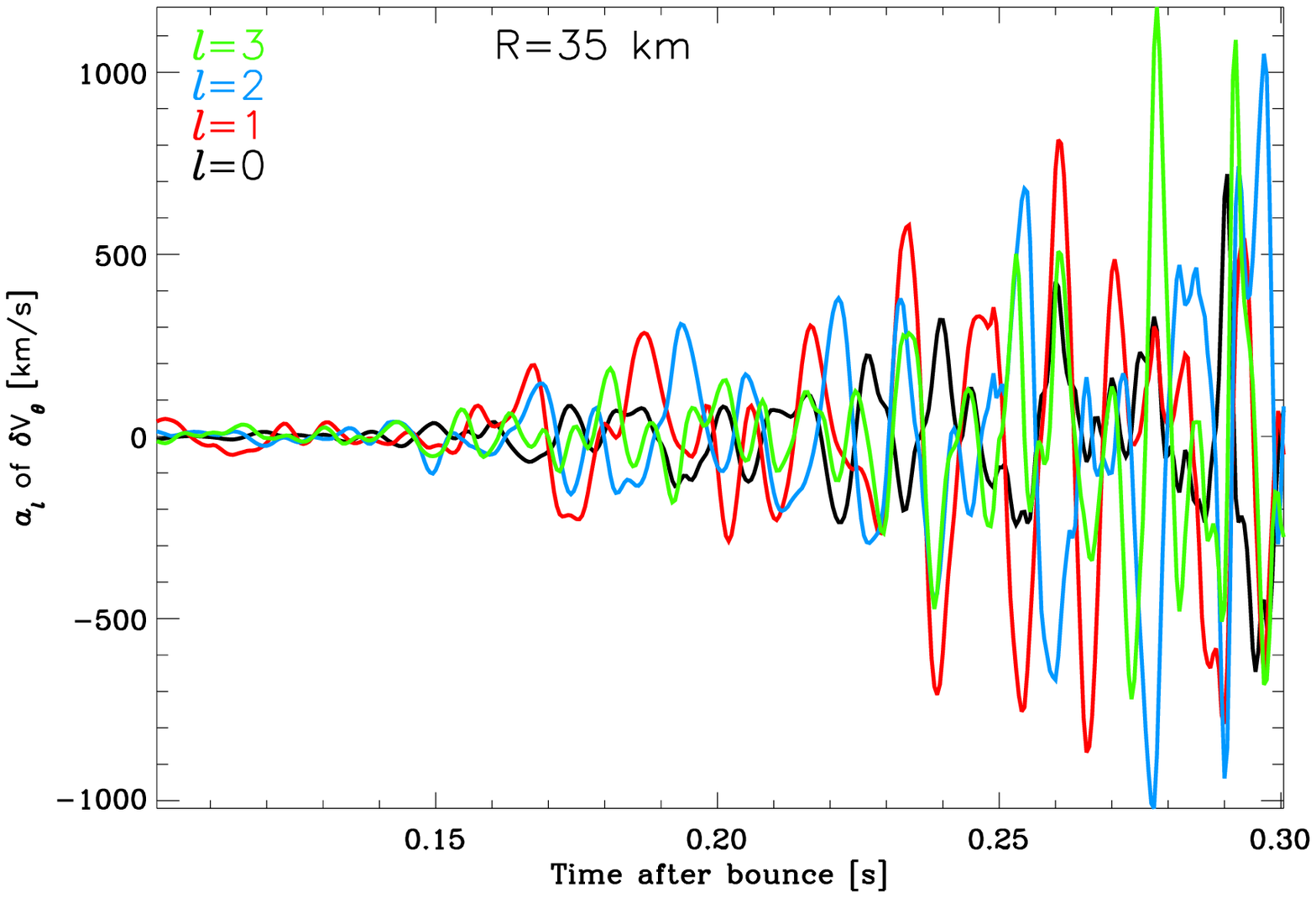}
\caption{
Time evolution after bounce of the spherical-harmonics coefficients for
modes $l=$0 (black), 1 (red), 2 (blue), and 3 (green) at a radius $R = 35$\,km, given by
$a_l =  2 \pi \int_0^{\pi} d\theta \sin\theta Y_l^0(\theta) \delta v_{\theta}(R,\theta)$.
}
\label{fig_ylm_vt}
\end{figure}

Geometrically, as was shown in Figs.~\ref{fig_entropy}-\ref{fig_density}, the radial extent of the cavity where
these waves exist is very confined, covering no more than 5--10\,km around 35\,km (at $\sim$200\,ms).
In the lateral direction, we again perform a Fourier transform of the latitudinal
velocity, this time subtracted from its angle average.
We show the resulting angular power spectrum in Fig.~\ref{fig_power_fft_vt_lat} with a maximum at a scale 
of 180$^{\circ}$ (the full range), and power down to $\sim$30$^{\circ}$.
There is essentially no power on smaller scales, implying a much larger extent of the waves in the lateral
direction than in the radial direction.
We also decompose such a latitudinal velocity field into spherical harmonics in order to extract the
coefficients for various l-modes.
We show the results in Fig.~\ref{fig_ylm_vt}, displaying the time-evolution
of the coefficients for $l$ up to 3, clearly revealing the dominance of $l=$1 and 2.

These characteristics are typical of gravity waves, whose horizontal $k_h$ and 
vertical $k_r$ wavenumbers are such that $k_h/k_r \ll 1$.
Moreover, the time frequency shown in (Fig.~\ref{fig_power_fft_vt}) corresponds very
well to the frequency of the large-scale overturning motions occurring in the layers above,
{\it i.e.}, $\nu_{\rm conv} \sim v_{\rm conv}/H_{\rm conv} \sim100$Hz, since typical 
velocities are of the order
of 5000\kms and $\Delta r$ between the PNS surface and the stalled shock is about 100\,km.
The behavior seen here confirms the analysis of Goldreich \& Kumar (1990) on (gravity-) wave excitation
by turbulent convection in a stratified atmosphere, with gravity waves having properties directly
controlled by the velocity, size, and recurrence of the turbulent eddies generating them.

\section{Summary and conclusions}
\label{conclusion}

In this paper, we have presented results from multi-dimensional radiation hydrodynamics 
simulations of protoneutron star (PNS) convection, providing support for the notion that large-scale 
overturn of core regions out to and above the neutrinosphere does not obtain, 
in agreement with studies by, e.g., Lattimer \& Mazurek (1981), Burrows \& Lattimer (1988), 
Keil et al. (1996), and Buras et al. (2005). 
Furthermore, the restricted convection is confined to a shell; no significant amount of 
neutrino energy from the diffusive inner regions makes it into the outer regions where 
neutrinos decouple from matter, thereby leaving the neutrino luminosity only 
weakly altered from the situation in which PNS convection does not occur.

We document our results by showing the spatial and time evolution for various thermodynamic and hydrodynamic
quantities, with 1) stills sampling the first 300\,ms past core bounce, 2) distribution functions, 3) time
series, and 4) frequency spectra.
In all simulations performed, convection occurs in two distinct regions that stay separate.
While convection in the outer region shows {\it negative} average radial-velocities, implying systematic
net accretion, it is associated in the inner region (radius less than 30\,km) with zero time- and angle-averaged 
velocities. In the interface region between the two convection zones lies a region where the radial velocity at
any time and along any direction is small. This effectively shelters 
the inner PNS from fierce convective motions occurring 
above 30\,km during these epochs. In this interface region, we identify the unambiguous 
presence of gravity waves, characterized with periods of 17\,ms and 8\,ms, 
latitudinal wavelengths corresponding to 30-180$^{\circ}$ (at 35\,km), and a radial extent of
no more than 10\,km.

The neutrinosphere radii, being highly energy dependent 50\,ms after bounce (from 20 to
$\ge$100\,km over the 1--200\,MeV range), become weakly energy-dependent 300\,ms after bounce (20 to 60\,km
over the same range). At 15\,MeV where the emergent $\nu_{\rm e}/{\bar{\nu}}_{\rm e}$ energy spectra peak at infinity, 
neutrinospheres shrink from $\sim$80\,km (50\,ms) down to $\sim$40\,km (300\,ms). 
This evolution results primarily from the mass accretion onto the cooling PNS, the cooling
and neutronization of the accreted material, and the concomitant steepening of the density gradient.

Importantly, the locations of the $\nu_e$ neutrinospheres 
are at all times beyond the sites of convection
occurring within the PNS, found here between 10 and 20\,km.  
As a result, there is no appreciable convective enhancement
in the $\nu_e$ neutrino luminosity. 
While energy is advected in the first $\sim$100\,ms to near the $\bar{\nu}_e$ and $\nu_{\mu}$
neutrinospheres and there is indeed a slight enhancement of as much as
$\sim$15\% and $\sim$30\%, respectively, in the total $\bar{\nu}_e$ and $\nu_{\mu}$ neutrino luminosities, 
after $\sim$100\,ms, this enhancement is choked off by the progressively increasing opacity barrier
between the PNS convection zone and all neutrinospheres.
Finally, we do not see overturning motions that could be interpreted as doubly-diffusive
instabilities in regions not unstable to Ledoux convection.

\acknowledgments

We acknowledge discussions with and help from
Rolf Walder, Jeremiah Murphy, Casey Meakin, Don Fisher, Youssif Alnashif, 
Moath Jarrah, Stan Woosley, and Thomas Janka. 
Importantly, we acknowledge support for this work
from the Scientific Discovery through Advanced Computing
(SciDAC) program of the DOE, grant number DE-FC02-01ER41184
and from the NSF under grant number AST-0504947.
E.L. thanks the Israel Science Foundation for support under grant \# 805/04,
and C.D.O. thanks the Albert-Einstein-Institut for providing CPU time on their
Peyote Linux cluster. We thank Jeff Fookson and Neal Lauver of the Steward Computer Support Group
for their invaluable help with the local Beowulf cluster and acknowledge
the use of the NERSC/LBNL/seaborg and ORNL/CCS/cheetah machines.
Movies and still frames associated with this work can be obtained upon request.

\appendix
\section{2D Flux-Limited Multi-Group Diffusion of Neutrinos}
\label{mgfld}

The MGFLD implementation of VULCAN/2D is fast and uses a vector
version of the flux limiter found in Bruenn (1985) (see also Walder et al. 2005).
Using the MGFLD variant of VULCAN/2D allows us to perform an extensive
study that encompasses the long-term evolution of many models. However, one
should bear in mind that MGFLD is only an approximation to full Boltzmann
transport and differences with the more exact treatment will emerge
in the neutrino semi-transparent and transparent regimes above the PNS
surface. Nevertheless, inside the neutrinosphere of the PNS the
two-dimensional MGFLD approach provides a very reasonable
representation of the multi-species, multi-group neutrino radiation fields.

 The evolution of the radiation field is described in the diffusion
approximation by a single (group-dependent) equation for the average 
intensity $J_g$ of energy group $g$ with neutrino energy $\varepsilon^g_{\nu}$:
\begin{equation}
  \dJgdt - div(D_g \nabla J_g) + \sigma^a_g J_g = S_g \, ,
\label{diff}
\end{equation}
where the diffusion coefficient is given by $D_g={1 \over 3\sigma_g}$ (and
then is flux-limited according to the recipe below), the total cross section 
is $\sigma_g$ (actually total inverse mean-free-path), and the absorption 
cross section is $\sigma^a_g$. The source term on the RHS of 
eq.~(\ref{diff}) is the emission rate of neutrinos of group $g$.
Note that eq.~(\ref{diff}) neglects inelastic scattering between
energy groups.

The finite difference approximation for eq.~(\ref{diff}) consists of
cell-centered
discretization of $J_g$. It is important to use cell-centered discretization
because the radiation field is strongly coupled to matter and the thermodynamic
matter variables are cell-centered in the hydrodynamical scheme. The finite
difference approximation of eq.~(\ref{diff}) is obtained by integrating the
equation over
a cell. Omitting group index and cell index one gets :
\begin{equation}
  V[\cdt (J^{n+1}-J^n) + \sigma_g^a J^{n+1}] +{\Sigma \dsi \cdot \BF_i^{n+1}} = V S_g \, .
\label{diffmgfld}
\end{equation}
Here $V$ is the volume of the cell, $\dsi$ is the face-centered vector
``$area_i \bf{n_i}$,"
$\bf{n_i}$ being the outer normal to face $i$. The fluxes $\BF_i$ at internal
faces
are the face-centered discretization of
\begin{equation}
 {\BF_i} = - D_i \nabla J^{n+1} \, ,
\label{flux}
\end{equation} 
where
\begin{equation}
 D_i = FL [{1 \over 3\sigma_i}]  \, .
\end{equation}
Our standard flux limiter, following Bruenn (1985) and Walder et al. (2005) is
\begin{equation}
 FL [D] = {D \over {1 + D |\nabla J| /J}}
\end{equation}
and approaches free streaming when $D$ exceeds the intensity scale height
$J/|\nabla J|$. The fluxes on the outer boundary of the system are defined
by free streaming outflow and not by the gradient of $J$.
Note that in eq.~(\ref{flux}) the fluxes are defined as face quantities, so that
they have exactly the same value for the two cells on both sides of that face.
The resulting scheme is therefore conservative by construction. In order to have
a stable scheme in the semi-transparent regions (large $D_g$) we center the
variables
in eq.~(\ref{diffmgfld}) implicitly. The fluxes, defined by the intensity at
the end
of the time step, couple adjacent cells and the final result is a set of
linear equations. The matrix of this system has the standard band structure
and we use a direct LU solver to solve the linear system. For a moderate grid
size the solution of a single linear system of that size does not overload
the CPU.

 In order to handle many groups, we have parallelized the code according
to groups. Each processor computes a few groups, usually less than 3, and
transfers the needed information to the other processors using standard MPI
routines. Since we do not split the grid between processors, the parallelization 
here is very simple. In fact, each processor performs the hydro step on
the entire grid. In order to avoid divergent evolution between different 
processors due to accumulation of machine round-off errors, we copy the 
grid variables of one chosen processor (processor 0) onto
those of the other processors, typically every thousand steps.

\end{document}